\documentclass[aps,prd,amsmath, showpacs, superscriptaddress,notitlepage]{revtex4}

\usepackage{graphicx}
\usepackage{hyperref}

\def\be{\begin{equation}}
\def\ee{\end{equation}}
\def\bea{\begin{eqnarray}}
\def\eea{\end{eqnarray}}

\begin{document}

\title{Self-gravitating scalar breathers with negative cosmological constant}

\author{Gyula Fodor}
\affiliation{Wigner Research Centre for Physics, RMKI, 1525 Budapest 114, P.O.Box 49, Hungary}
\affiliation{LUTH, CNRS-UMR 8102, Observatoire de Paris-Meudon, place Jules Janssen, 92195 Meudon Cedex, France}\author{P\'eter Forg\'acs}
\affiliation{Wigner Research Centre for Physics, RMKI, 1525 Budapest 114, P.O.Box 49, Hungary}
\affiliation{LMPT, CNRS-UMR 6083, Universit\'e de Tours, Parc de Grandmont, 37200 Tours, France}
\author{Philippe Grandcl\'ement}
\affiliation{LUTH, CNRS-UMR 8102, Observatoire de Paris-Meudon, place Jules Janssen, 92195 Meudon Cedex, France}


\begin{abstract}
Breather-type (time-periodic and spatially localized) solutions with spherical symmetry are investigated in a massless scalar field theory coupled to Einstein's gravity with cosmological constant in $d$ spatial dimensions imposing anti de Sitter (AdS) asymptotics on space-time. Using a code constructed with the Kadath library that enables the use of spectral methods, the phase space of breather
solutions is explored in detail for $d=3$ and $d=4$. It is found that there are discrete families of solutions indexed by an integer and by their frequency. Using a time evolution code these AdS breathers are found to be stable for up to a critical central density, in analogy to boson stars. Using an analytical perturbative expansion small amplitude breathers are worked out for arbitrary dimensions $d$.
\end{abstract}

\maketitle

\today

\section{Introduction} \label{sec:intro}
There is a considerable interest to understand the gravitational dynamics of asymptotically anti de Sitter (AdS) space-times, stimulated to a large extent by the AdS/CFT correspondence, but independently of that, the problem is of sufficient inherent interest on its own \cite{maliborski3,deppe,dimitrakopoulos,kim,craps14,craps,buchel-15,Farahi14,friedrich14}. 
In the ground breaking work \cite{bizon-11}, the time-evolution of a free, massless scalar field coupled to Einstein's gravity has been investigated in $3$ spatial dimensions in asymptotically AdS space-times, with the result that from a large class of smooth initial data black holes form, indicating that AdS is unstable. 
By now a large body of rather convincing numerical and perturbative evidence has accumulated
that asymptotically AdS space-times (having the same conformal boundary that AdS) are in fact unstable with respect to black hole formation 
\cite{jalmuzna-11,buchel-12,dias1,dias2,garfinkle12,maliborski12,liebling13,oliveira13,bizon14,
balasubramanian14,shengyang15,okawa15,evnin15}. 
The fact that arbitrarily small perturbations of an AdS space-time lead generically to black hole formation is in sharp contradistinction to the Minkowski or de Sitter cases which are stable. The instability uncovered by Ref.\ \cite{bizon-11} manifest itself by the fact that more and more energy gets concentrated in the same spatial region by weak turbulence. The existence of such an instability is clearly related to the peculiar causality structure of asymptotically AdS space-times, which are not globally hyperbolic, ie.\ there is no Cauchy hypersurface in them.
As a consequence, the prescription of initial data on a space-like hypersurface in the usual way is not sufficient to uniquely determine its time evolution. In contradistinction to asymptotically flat or de Sitter space-times  one also has to specify suitable boundary conditions on the time-like conformal boundary of the asymptotically AdS space-time at (null and spatial) infinity.

In Ref.\ \cite{maliborski1} it has been found that the same system, a minimally coupled, free, massless scalar field coupled to gravity, as in Ref.\ \cite{bizon-11} admits spherically symmetric, time-periodic regular solutions, whose mass is finite. Such localized, time-periodic solutions shall be referred to as AdS breathers. In Ref.\ \cite{maliborski1} AdS breathers have been constructed both by perturbative expansion and by direct numerical calculation, although, possibly because of the letter form of the paper, concrete results have been presented only for $4+1$ spacetime dimensions. Both the perturbative and the direct numerical methods in \cite{maliborski1} are based on the expansion in terms of the eigenfunctions of the linearized problem. In case of even spatial dimensions, to each order in the perturbative expansion, all metric and scalar field functions can be expressed as a sum of finite number of these terms. This does not hold for odd spatial dimensions, making the small-amplitude expansion formalism in \cite{maliborski1} not applicable in practice in that case. Similarly, the presence of slowly decreasing coefficients in terms of the linearized solutions is likely to make the direct numerical method to determine the breather in \cite{maliborski1} impractical for even space dimensions.

Breather-type solutions are of obvious interest and it is a particularly important question to clarify their main properties, in particular their stability. In this paper we present a numerical construction of spherically symmetric AdS breathers using the Kadath library \cite{Grand09,Kadath}, on the one hand, and compute them analytically by a perturbative expansion in the amplitude, on the other hand. Importantly, we also investigate the stability of the AdS breathers using a time evolution code. We concentrate on the $d=3$ and $d=4$ dimensional cases, but our methods work in arbitrary dimensions.

Our numerical construction of the AdS breathers is based on deforming the well known linearized AdS breathers \cite{avisisham} having eigenfrequencies $\omega_n=d+2n$ where $n=0,1,\ldots$ is the node-number, whereby the amplitude of the solution becomes frequency dependent. We have mostly constructed breathers corresponding to the nodeless mode, $n=0$. 
In Ref.\ \cite{HorSan} gravitational geons -- localized, asymptotically AdS vacuum solutions of the Einstein equations with cosmological constant -- have been calculated in $3+1$ dimensions. Nevertheless, to the best of our knowledge, scalar field AdS breathers in $3+1$ dimensions have not been constructed numerically in the literature before.
In case of $4+1$ dimensional spacetimes, careful comparison of our solutions with those of Ref.\ \cite{maliborski1} shows very good agreement, providing rather nontrivial checks on the validity of both approaches. We note that even for $4+1$ dimensions our approach is able to provide significantly larger amplitude solutions than those presented in \cite{maliborski1}. The likely reason for this is the choice of the parameter $\varepsilon$ in \cite{maliborski1}, which reaches a maximal value already before the amplitude of breather grows to the value where the mass is maximal. 

Importantly, we have found that AdS breathers whose frequencies lie between $\omega_s$ and $d$, where $\omega_s\approx 2.253$ for $d=3$ and $\omega_s\approx 3.548$ for $d=4$, are stable. The change of stability occurs where the mass of the AdS breather is maximal, in analogy to what happens for boson stars \cite{Jetzer,SchunkMielke,buchel-13}.

We also find a rather complicated resonance structure of the nodeless breathers both in $d=3$ and $d=4$, where the amplitude of certain higher Fourier modes of the scalar field increases abruptly. We have observed several such peaks, although here we only present numerical evidence for those which are easiest to find numerically. The detailed study of these resonances would require further extremely high precision numerical runs. However, the importance of their existence is reduced by the fact that apparently they all lie in the high amplitude unstable domain. The smallest frequency where our numerical code converges is around $\omega\approx 2.1$ in $d=3$ and  $\omega\approx 3.5$ in $d=4$. The existence of a minimal frequency for each $d$ and node number, $n$ is expected.

The paper is organized in the following way. In Sec.~\ref{sec:periodic} we present the spectral methods used for the construction of time-periodic solutions. First, we compute the linearized solutions of the problem, and establish the agreement with the analytical results. These solutions are used as initial guess for the iteration procedure for obtaining the solutions of the full nonlinear problem in $d=3$ and $d=4$. For $d=4$ spatial dimensions we find very good agreement between our results and those of \cite{maliborski1}. In Sec.~\ref{s:evolution} the numerical time evolution code is presented. We show that numerically constructed AdS breathers are stable under time evolution as long as their frequencies are above $\omega_s$, and those with lower frequencies are unstable, and form black holes. In Sec.~\ref{sec:smallampl} a small-amplitude expansion procedure is presented for arbitrary dimensions. We have performed the expansion up to fourth order, and excellent agreement with the numerical results have been found for as high values of the expansion parameter as $\varepsilon\approx 1$. This strongly suggest that our perturbative expansion has a finite convergence radius.

\section{Construction of periodic solutions by spectral method} \label{sec:periodic}

\subsection{Field equations with a massless scalar field}\label{ss:field}

We consider $d+1$ dimensional Einstein's equations with a negative cosmological constant $\Lambda$,
\begin{equation}
G_{\mu\nu}+\Lambda g_{\mu\nu}=8\pi G T_{\mu\nu} \ , \label{eingen}
\end{equation}
where the stress-energy tensor is provided by a minimally coupled zero mass real Klein-Gordon scalar field $\phi$,
 \begin{equation}
T_{\mu\nu}=\phi_{,\mu}\phi_{,\nu}-\frac{1}{2}g_{\mu\nu}\phi_{,\alpha}\phi^{,\alpha} \ ,
\end{equation}
where all the Greek indices are $d+1$-dimensional. The divergence of the Einstein's equations yields the wave equation
\begin{equation}
g^{\mu\nu} \nabla_{\mu} \nabla_{\nu} \phi = 0 \ . \label{waveeqgen}
\end{equation}
Since for $d=1$ spatial dimensions both the Einstein tensor and the energy momentum tensor is traceless, equation \eqref{eingen} can only be satisfied for zero cosmological constant. Hence in the following we assume $d\geq 2$.

\subsection{Choice of coordinates}

The spectral code used in this section works in coordinates such that $G=1$. This is to be contrasted with the evolution code of Sec. \ref{s:evolution} which employed $8\pi G = d-1$. The reader should keep that in mind when comparing results coming from various sources.

There are many possible choices of coordinates for the AdS spacetime, and it is also the case for spacetimes that are only asymptotically AdS. For the spectral methods applied in this section a natural choice is to apply isotropic coordinates. In those coordinates the $d+1$ dimensional AdS metric reads
\be
\label{e:met}
{\rm d}s^2 = - \left(\frac{1+\rho^2}{1-\rho^2}\right)^2 {\rm d}t^2 + \left(\frac{2}{1-\rho^2}\right)^2 \left({\rm d}r^2 + r^2 {\rm d}\Omega^2_{d-1}\right) ,
\ee
where $\rho$ is defined as $\rho=r/L$, and $\mathrm{d}\Omega_{d-1}^2$ is the metric on the $d-1$ dimensional unit sphere. The radius $r$ goes from $0$ to $L$, so that $\rho \in\left[0, 1\right]$. The maximum radius $L$ is related to the cosmological constant by
\be
\Lambda = - \frac{d(d-1)}{2L^2} \ . \label{e:lambda}
\ee
We note that rescaling the time and radial coordinates one can always set $L=1$, which we will employ in actual numerical calculations.

In the language of the $d+1$ formalism, \eqref{e:met} means that the spacetime is simply described by a lapse function $N$, a conformal factor $\Psi$, and no shift vector, so that
\be
\label{e:metdp1}
{\rm d}s^2 = - N^2 {\rm d}t^2 + \Psi^4 \left({\rm d}r^2 + r^2 {\rm d}\Omega^2_{d-1}\right).
\ee
A simple identification yields the expression of the metric potential in the AdS case:
\be
\label{e:pureads}
N =  \frac{1+\rho^2}{1-\rho^2} \quad ; \quad
\Psi = \left(\frac{2}{1-\rho^2}\right)^{1/2}.
\ee
Close to $r\rightarrow L$, the two metric fields diverge. More precisely, they behave like
\be
\label{e:asymptot}
N = \frac{L}{\epsilon} \quad ; \quad
\Psi = \left(\frac{L}{\epsilon}\right)^{1/2} ,
\ee
where $\epsilon = L - r$.

When one deals with spacetimes that are only asymptotically AdS, the expressions (\ref{e:pureads}) no longer hold. The metric can still be described by the form (\ref{e:metdp1}). As we are interested in spherically symmetric and periodic solutions, the fields $N$ and $\Psi$ depend only on $r$ and $t$. The proper geometry at $\rho=1$ is recovered by imposing the AdS asymptotic behaviors (\ref{e:asymptot}).

Various $d+1$ projections of the stress-energy tensor will appear in the decomposition of the field equations. The energy density is given by
\begin{equation}\label{e:endens}
E = \frac{1}{2N^2} \left(\phi_{,t}\right)^2 + \frac{1}{2\Psi^4} \left(\phi_{,r}\right)^2 \ ,
\end{equation}
and the trace of the spatial part of the stress-energy tensor is
\begin{equation}\label{e:trstress}
S = \frac{d}{2N^2} \left(\phi_{,t}\right)^2 - \frac{d-2}{2\Psi^4} \left(\phi_{,r}\right)^2 \ .
\end{equation}

\subsection{\texorpdfstring{\lowercase{d}}{d}+1 equations}

Even if there is no shift vector, due to the fact that the fields depend on $t$, there is an extrinsic curvature tensor. Its expression comes from the kinematic definition of the extrinsic curvature tensor, which simply reads: $\partial_t {\gamma_{ij}} = -2 N K_{ij}$, where  $\gamma_{ij} = \Psi^4 f_{ij}$ is the d-dimensional spatial metric and $f_{ij}$ the d-dimensional flat spatial metric. It follows that
\be
K_{ij} = -\frac{2}{N} \left(\frac{\Psi_{,t}}{\Psi}\right) \Psi^4 f_{ij} \ .
\ee
The Hamiltonian constraint equation, with a cosmological constant, reads $R + K^2 - K_{ij} K^{ij} - 2 \Lambda = 16\pi G E$, where $E$ is the $d+1$ energy density given by \eqref{e:endens}. It leads to
\be\label{e:eqpsi}
\Psi_{,rr} + \left(d-1\right) \frac{\Psi_{,r}}{r} + \left(d-3\right)
\frac{\left(\Psi_{,r}\right)^2}{\Psi} - \frac{d}{N^2} \Psi^3 \left(\Psi_{,t}\right)^2 =
\frac{d\Psi^5}{4 L^2} - \frac{4\pi G}{d-1} \Psi^5 E \ .
\ee
The second equation that we solve is obtained by taking the trace of the evolution equation for $K_{ij}$. For zero shift this takes the general form
\begin{equation}
\gamma^{ij}\partial_t K_{ij}=-D^i D_i N-N K_{ij} K^{ij}
-\frac{2N}{d-1}\Lambda
+ 8\pi G N \frac{(d-2)E+S}{d-1} \ ,
\end{equation}
and using our variables it gives the following result
\begin{align}\label{e:eqlapse}
N_{,rr}& + \left(d-1\right) \frac{N_{,r}}{r}
+ 2 \left(d-2\right) N_{,r} \frac{\Psi_{,r}}{\Psi} =\\
&\Psi^4 \left[\frac{2d}{N} \frac{\Psi_{,tt}}{\Psi}
- \frac{2d}{N} \frac{N_{,t}}{N} \frac{\Psi_{,t}}{\Psi}
+ \frac{2d}{N} \left(\frac{\Psi_{,t}}{\Psi}\right)^2\right]
+ N \Psi^4 \frac{d}{L^2}
+ 8\pi G N \Psi^4 \frac{(d-2)E+S}{d-1}  \ , \nonumber
\end{align}
where $S$ is the trace of the spatial stress-energy tensor given by \eqref{e:trstress}. The wave equation \eqref{waveeqgen} reads
\be
\label{e:field}
-\frac{1}{N^2} \left[\phi_{,tt} + 2d \frac{\Psi_{,t}}{\Psi} \phi_{,t} - \frac{N_{,t}}{N} \phi_{,t}\right]
+ \frac{1}{\Psi^4} \left[\phi_{,rr} + \left(d-1\right) \frac{\phi_{,r}}{r}+ \frac{N_{,r}}{N} \phi_{,r} + 2 \left(d-2\right) \frac{\Psi_{,r}}{\Psi} \phi_{,r}\right] = 0 \ .
\ee

\subsection{Equations in the region close to infinity}\label{ss:far}

We aim at solving the equations numerically, using spectral methods (see Sec. \ref{ss:numerics}). In order to maintain the precision of spectral methods, it is then desirable to work with ${\mathcal C}^\infty$ functions. This is not the case of $N$ and $\Psi$ that diverge at $r=L$. Motivated by the asymptotics (\ref{e:asymptot}), in the region close to $L$, the following auxiliary variables are used
\be
n = \frac{1}{N} \quad ; \quad f=\frac{1}{\Psi^2}.
\ee
Both $n$ and $f$ vanish linearly when $r\rightarrow L$ and are well described by our spectral expansion. Equations (\ref{e:eqpsi}), (\ref{e:eqlapse}) and
(\ref{e:field}) can be rewritten in terms of $n$ and $f$ and multiplied by the appropriate powers to get rid of all diverging quantities. They give rise (respectively) to the following set of equations:
\bea
\label{e:eqf}
f^3 f_{,rr} +  f^3 \frac{d-1}{r} f_{,r}- \frac{d}{2} f^2 \left(f_{,r}\right)^2
+ \frac{d}{2} n^2 \left(f_{,t}\right)^2 + \frac{d}{2} f^2 \frac{1}{L^2}
- \frac{8\pi G}{d-1}f^2  E &=& 0 \ ,\\
\label{e:eqn}
-nf^4 n_{,rr} - n f^4 \frac{d-1}{r}n_{,r} + 2 f^4 \left(n_{,r}\right)^2
+ \left(d-2\right)nf^3 n_{,r} f_{,r} + d n^4 f f_{,tt} - 2d n^4 \left(f_{,t}\right)^2 + &&
\nonumber \\
d n^3 f n_{,t} f_{,t} - \frac{dn^2f^2}{L^2}
- 8\pi G n^2f^2  \frac{\left(d-2\right)E+S}{d-1}&=& 0 \ , \\
\label{e:phifar}
- n^3 f \phi_{,tt} + d n^3 f_{,t} \phi_{,t} - n^2 f n_{,t} \phi_{,t} + n f^3 \phi_{,rr} + n f^3 \frac{d-1}{r} \phi_{,r} - f^3 n_{,r} \phi_{,r} - \left(d-2\right)n f^2 f_{,r} \phi_{,r} &=& 0 \ .
\eea

The numerical boundary conditions at $r=L$ are given by looking at the terms that vanish the less rapidly in Eqs. (\ref{e:eqf}) and (\ref{e:eqn}), respectively. It follows that at infinity
\bea
\label{e:nfar}
2 n_{,r}^2 + \left(d-2\right) n_{,r} f_{,r} &=& \frac{d}{L^2} \ , \\
\label{e:ffar}
f_{,r} &=& -\frac{1}{L} \ .
\eea
Those conditions are supplemented by the fact that the field vanishes at infinity: $\phi\left(r=L\right)=0$. It is also possible, from the numerical solutions, to verify that the conditions (\ref{e:nfar}) and (\ref{e:ffar}) do lead to asymptotically AdS solutions (i.e. solutions for which $f=0$ and $n=0$ when $r=L$).

\subsection{Numerical methods}\label{ss:numerics}

The numerical setting used in this paper is very similar to the one employed in our previous work \cite{FodorFG14}. The Kadath library \cite{Grand09, Kadath} is used to numerically solve the equations at hand. The library enables the use of spectral methods. Here a two-dimensional setting is used, for the fields do depend only on $r$ and $t$. The radial coordinate ranges from $0$ to $L$ and several numerical domains (typically 4) are used. In each domain, there is an affine law between the true radial coordinate $r$ and the numerical one $r^\star$. It ensures that $r^\star \in\left[-1,1\right]$ ($\left[0, 1\right]$ near the origin). Spectral expansion is performed with respect to $r^\star$ and Chebyshev polynomials are used (even ones only near the origin). The influence of the number of radial domains on the precision is difficult to assert beforehand. However there is no real necessity to measure this effect. What is more meaningful is to verify that the error converges with the number of spectral coefficients fast enough that the required accuracy can be achieved in practice. This is the object of Sec. \ref{ss:precision}.

As far as time is concerned, one aims at finding periodic solutions so that the fields are expanded onto trigonometrical functions. More precisely, it can be shown that the scalar field $\phi$ can be expanded onto odd cosines and the metric fields onto even ones. Only one temporal domain is used where the numerical time $t^\star=\omega t$ ranges from $0$ to $\pi/2$.

The mathematical system to be solved consists of Eqs (\ref{e:eqpsi}), (\ref{e:eqlapse}) and (\ref{e:field}). In the numerical domain that contains $r=L$ they are replaced by Eqs. (\ref{e:eqf}), (\ref{e:eqn}) and (\ref{e:phifar}). Regularity of the fields is ensured at $r=0$ and the appropriate boundary conditions at infinity are discussed in Sec. \ref{ss:far} and given by Eqs. (\ref{e:nfar}) and (\ref{e:ffar}) and by demanding that the field vanishes at $r=L$.

\begin{figure}[!hbtp]
\includegraphics[width=8cm]{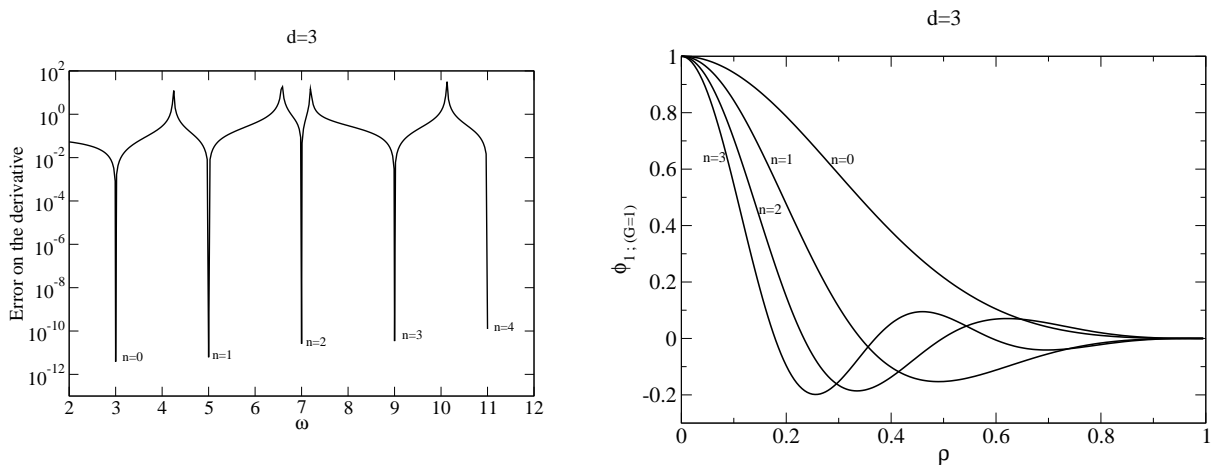}
\includegraphics[width=8cm]{prof_linear.eps}
\caption{\label{f:lin} The first panel shows the error on the radial derivative as a function of $\omega$, for $d=3$. The location of the zeros correspond to the true solutions of the small amplitude expansion. The second panel shows the different solutions with increasing number of nodes, i.e. for increasing values of $\omega$.}
\end{figure}

The resulting system is solved using a Newton-Raphson iteration. This requires the use of an initial guess configuration. One possible method consists in using a result coming from a small amplitude expansion. Given the fact that the stress-energy tensor is quadratic in $\phi$, at lowest order one can fix the geometry to AdS (Eq. (\ref{e:met})) and solve only for the wave equation. In order to avoid the trivial solution $\phi = 0$, we use a similar technique than the one used for the study of massive scalar fields on fixed adS background. We impose that the first harmonic takes a given value at an intermediate radius $r_{\rm mid}$ : $\Psi\left(r_{\rm mid}\right) = 0.1 * \cos \left(\omega t\right)$.
This condition is used as an outer boundary condition for $r<r_{\rm mid}$ and as an inner boundary condition for $r>r_{\rm mid}$. Doing so one finds a non-zero solution but which radial derivative is, in general, not continuous at $r=r_{\rm mid}$. Only for some discrete values of $\omega$ is the solution ${\mathcal C}^\infty$, representing a true solution. Figure \ref{f:lin} shows some results for the linear solution in the case $d=3$. In this case one recovers the fact that the (angular) frequencies of the small amplitude solutions are $\omega_n = 3 + 2n$ and that the solution with $\omega_n$ has $n$ nodes. The error on the continuity is of the order of $10^{-10}$, being typically the overall precision of the code (due to finite number of coefficients and round-off errors).

Small amplitude configurations are used as an initial guess for computing non-linear solutions. One works at a frequency close to the linear one: $\omega = \omega_n - \Delta_\omega$ and one scales the linear solution so that $\phi\left(r=0, t=0\right) = \delta_\phi$. With an appropriate choice of $\Delta_\omega$ and $\delta_\phi$, it is usually possible to achieve convergence to a non-vanishing solution of the full system. Once a first configuration is found, a sequence can be constructed by slowly varying $\omega$.

\subsection{Assertion of the precision}\label{ss:precision}

\begin{figure}[!hbtp]
\includegraphics[width=8cm,clip]{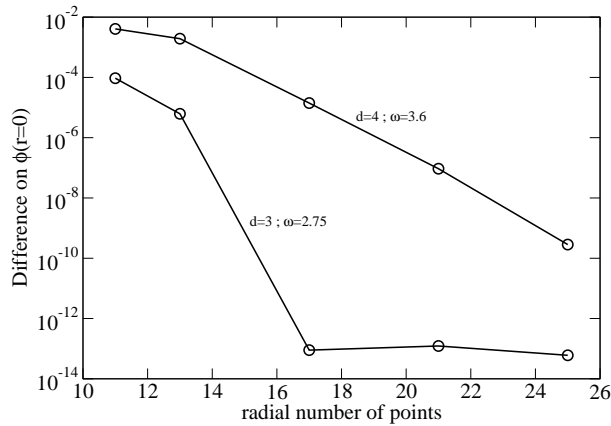}
\caption{\label{f:conv} Difference between the result for $\left(N_r, N_t\right) = \left(33, 17\right)$ and various lower resolution results. The y-axis denotes difference in the value of the scalar field at the origin and the x-axis the number of radial coefficients $N_r$.}
\end{figure}

A first check is concerned with the spectral convergence of the solutions. By this one means a convergence faster than any power-law, as the resolution increases. In practice an exponential convergence is often observed. This is one of the most striking advantage of using spectral methods. One resolution is a set of two integers $\left(N_r, N_t\right)$, being respectively the number of coefficients used for the radial variable (in each radial domain) and for the time. For largely empirical reasons, the following resolutions are considered (keeping roughly $N_r \approx 2 N_t$) : $\left(11, 7\right)$, $\left(13, 9\right)$, $\left(17, 11\right)$, $\left(21, 13\right)$, $\left(25, 15\right)$ and $\left(33, 17\right)$. Figure  \ref{f:conv} shows the difference between the value of $\phi\left(r=0\right)$ obtained at the highest resolution and the one for various lower resolutions (i.e. in a sense considering the value for $\left(33, 17\right)$ as being the "true" value). Results are shown for both $d=3$ and $d=4$. In the case $d=4$ the spectral convergence is easily seen. For the $d=3$ the convergence is actually so fast that the round-off error of $10^{-13}$ is rapidly reached. It is true that convergence of the result does not always mean precision but it is usually a rather good indicator. In the following, the various results are obtained for the highest resolution, that is for $\left(N_r, N_t\right) = \left(33, 17\right)$.

In order to further assert the validity of our numerical code, results are compared to the ones obtained, for $d=4$ in \cite{maliborski1}. It is a strong test because the methods and coordinates are very different. It is indeed so much the case that even just comparing the results is not at all straightforward. First one needs to compute, from our solution, the value of the parameter $\varepsilon$ used in \cite{maliborski1}. It can be shown to be the scalar product of $\phi\left(t=0\right)$ and the lowest order linear solution, which is $e_0$ in the language of \cite{maliborski1}. Using their equation (11), one can see that $e_0 \left(x\right) = 2 \sqrt{6} \cos^4 x$. The scalar product in question is the one used in \cite{maliborski1}, that is the one with a weight function $\tan^3 x$.  It follows then that
\be
\label{e:epsilon}
\varepsilon = 2 \sqrt{6} \int_0^{\pi/2} \phi\left(r,t=0\right) \cos x \sin^3 x{\rm d}x.
\ee
We note that this $\varepsilon$ is different from the small-amplitude expansion parameter $\varepsilon$ used in Sec.~\ref{sec:smallampl} of our paper.

In order to compute the integral appearing in Eq.~(\ref{e:epsilon}), one needs to express the radial coordinate $r$ as a function of the coordinate $x$ used in \cite{maliborski1}. A direct comparison of the line elements, for $t=0$, shows that $r$ must be the solution of $\Psi^2\left(r\right)r = L\tan x$, which can be solved easily using, for instance, a secant method. One must also recall that the spectral numerical code in this paper uses units such that $G=1$ whereas \cite{maliborski1} uses $8\pi G = 3$ so that $\varepsilon$ must be scaled by an additional factor
$\sqrt{8 \pi/3}$.

The last complication comes from the fact that we measure the frequency $\omega$ at infinity whereas the $\Omega$ used in \cite{maliborski1} is defined at the origin. One can show that the two relates by $\Omega = \omega / N_0 \left(r=0\right)$ where $N_0$ is the constant part of our lapse function (remember that $N = \sum_j N_j \cos\left(2j \omega t\right)$ ; see Sec. \ref{ss:numerics}).

\begin{figure}[!hbtp]
\includegraphics[width=8cm,clip]{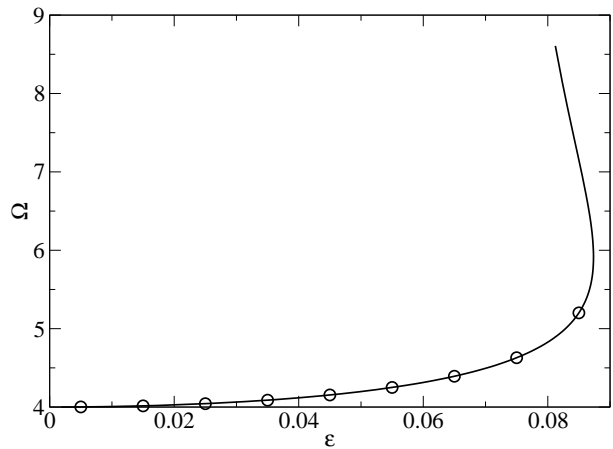}
\caption{\label{f:comppolish} Values of the central frequency $\Omega$ as a function of the expansion parameter $\varepsilon$ used in \cite{maliborski1}. The solid line denotes the results from this paper and the circles the data taken from Table II of \cite{maliborski1}.}
\end{figure}

Figure \ref{f:comppolish} shows the comparison between our results and the ones from \cite{maliborski1}. More precisely, it shows the value of $\Omega$, as a function of $\varepsilon$ for both codes. The agreement is so good that the results are indistinguishable by eye and a relative difference of order $10^{-4}$ is measured on the whole range of $\varepsilon$. We view this as being sufficiently small to confirm the validity both works. More precision (i.e. smaller error) could probably be achieved by increasing the precision of the comparison itself (for instance when computing the integral (\ref{e:epsilon})).

\begin{figure}[!hbtp]
\includegraphics[width=8cm,clip]{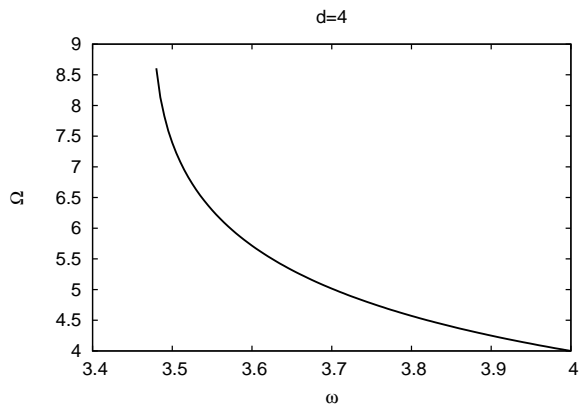}
\caption{\label{f:omom} Oscillation frequency $\Omega$ measured by a central observer, as function of the frequency $\omega$ observed by a faraway observer, for $d=4$ spatial dimensions. For small amplitude states both frequencies approach the value $4$.}
\end{figure}

A striking feature of Fig. \ref{f:comppolish} is the fact that the parameter $\varepsilon$ has a maximum value of about $0.087$. This explains why the code used in \cite{maliborski1} fails to converge for $\varepsilon > 0.085$. Nevertheless, the sequence of solution continues to exist, with increasing values of $\Omega$. This indicates that the variable $\varepsilon$ is probably not the best choice to parametrize the various solutions. In our case, the parameter $\omega$ is used instead and it seems to lead to an easier exploration of the parameter space, as illustrated by the results shown in Sec.~\ref{ss:results}.
In order to facilitate the comparison with the results in \cite{maliborski1} we plot the dependence of the central frequency $\Omega$ on the asymptotic frequency $\omega$ on Fig.~\ref{f:omom}

\subsection{Numerical results}\label{ss:results}

\begin{figure}[!hbtp]
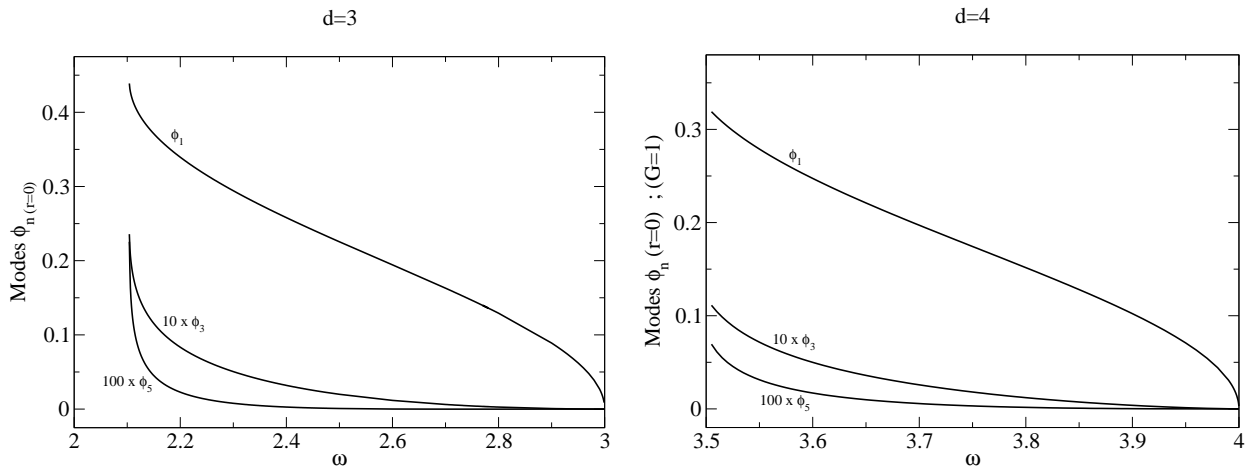

\includegraphics[width=8cm]{ori_d3.eps}
\quad
\includegraphics[width=8cm]{ori_d4.eps}
\caption{\label{f:ori} Values of the various modes $\phi_n$ at the origin, for $d=3$ (first panel) and $d=4$ (second panel), as a function of $\omega$. The modes are scaled for plotting convenience.}
\end{figure}

In this section, some numerical results are presented for $d=3$ and $d=4$. We consider only configurations that connect to the nodeless linear ones. First, Fig. \ref{f:ori} shows the central value of the first three modes
$\phi_n$ of the scalar field (remember that only the odd modes are non-zero). As expected all the various modes go to zero when one approaches the frequency of the linear solution (which is $\omega=d$ in this case). Even in the case $d=4$, the curves are relatively monotonic and it confirms the fact that the behavior exhibited in Fig. \ref{f:comppolish} comes solely from the use of the variable $\varepsilon$.

\begin{figure}[!hbtp]
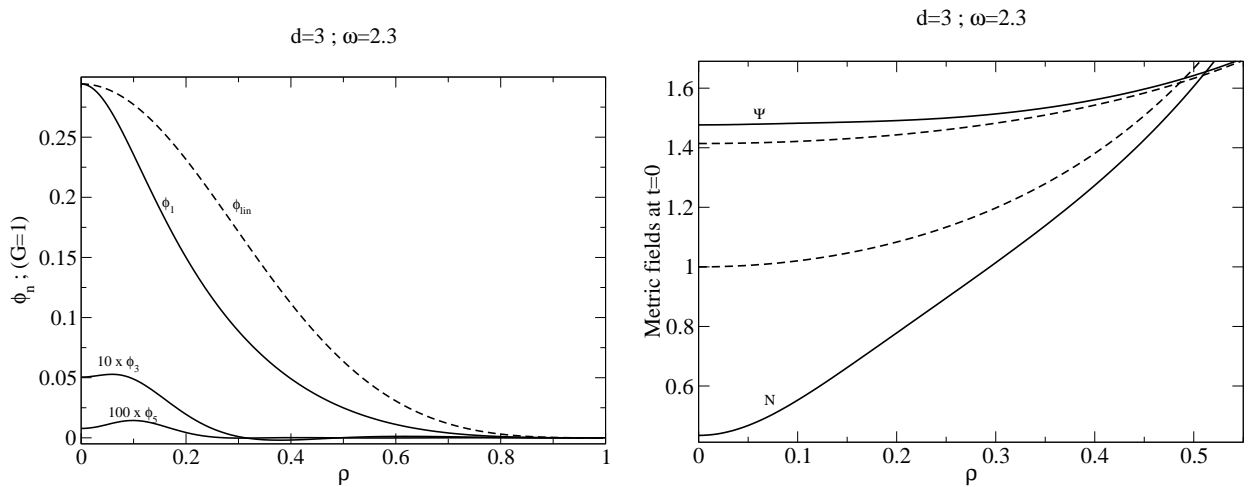

\includegraphics[width=8cm]{phi_2.3.eps}
\quad
\includegraphics[width=8cm]{metric_2.3.eps}
\caption{\label{f:data_2.3} Solution for $d=3$ and $\omega=2.3$. The first panel shows the first modes of the scalar field and the second one the metric fields at $t=0$. The dashed lines are the linear solution (first panel) and the solution for the exact AdS spacetime (second panel).}
\end{figure}

Figure \ref{f:data_2.3} shows the spatial structure of the solution for one particular example, corresponding to $d=3$ and $\omega = 2.3$. The first panel shows the various modes of the scalar field $\phi_n$ and the second one the values of the metric function $N$ and $\Psi$, at $t=0$. The dashed curves represent the linear solution $\phi_{\rm lin}$ (first panel) and the metric fields corresponding to pure AdS spacetime (second panel).
Although with our code the solutions stop at $2.104$ for $d=3$, and at $3.48$ for $d=4$, we do not think that the solutions simply disappear for such values of the frequency, rather we expect the families of solutions to continue. It seems plausible, in analogy with boson stars, that close to these points the solution curves turn back, and the new branch continues for increasing values of the frequency.

\begin{figure}[!hbtp]
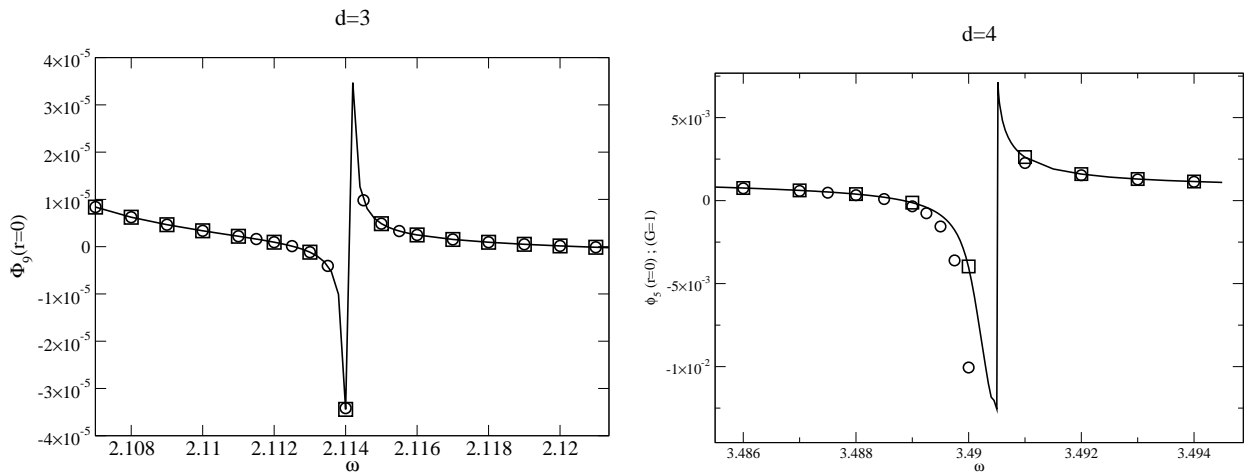

\includegraphics[width=8cm]{reson_d3.eps}
\quad
\includegraphics[width=8cm]{reson_d4.eps}
\caption{\label{f:reson} Mode $\phi_n$ at the origin, around two resonances found for $n=9$ and $d=3$  (first panel) and $n=5$ ad $d=4$ (second panel). The solid lines denote the high resolution results $\left(N_r, N_t\right) = \left(33, 17\right)$, the squares the middle resolution ones $\left(N_r, N_t\right) = \left(25, 15\right)$, and the circles the lower resolution ones $\left(N_r, N_t\right) = \left(21, 13\right)$.}
\end{figure}

One of the most interesting findings of our previous work \cite{FodorFG14} on fields on fixed AdS background was the existence of many resonances (see Sec.~V-C of \cite{FodorFG14}).
The solutions obtained in this paper are asymptotically AdS and we also found a number of resonances in analogy with the fixed AdS background ones.
The resonances are, however, relatively difficult to find because they are very narrow and noticeable only in rather high order modes. Two such resonances are shown in Fig.~\ref{f:reson}, for $d=3$ (first panel) and $d=4$ (second panel). In order to check that they are real Fig. \ref{f:reson} also shows some results obtained with two lower resolutions (the circles correspond to $\left(N_r, N_t\right) = \left(21, 13\right)$ and the squares to  $\left(25, 15\right)$). For $d=3$ the various resolutions are indistinguishable and a clear convergence is seen in the case $d=4$. This makes us very confident that the observed resonances are genuine and not numerical artifacts.

We have not seen major differences in the resonances between the cases $d=3$ and $d=4$, in terms of shape (this is to be contrasted with the ones observed in \cite{FodorFG14} which shape could vary from one case to the other). On Fig.~\ref{f:reson} the ninth mode has been plotted for $d=3$ and the fifth one for $d$=4. They correspond to the modes where the the resonances are the easiest to see: they are also visible, although much less noticeable, in the other modes. Let us state that in theory the resonance is expected to be present in every mode due to the coupling between them. However, this would require a very precise fine-tuning of the frequency to be seen clearly.

In \cite{FodorFG14} a relatively simple explanation for the appearance of the resonances was given: basically, one of the modes became dominant and  was approaching one of the small amplitude solutions (see the second panel of Fig. 9 in \cite{FodorFG14}). Unfortunately no such simple interpretation is found to hold in the case at hand. It might come from the fact that the resonances correspond to cases far from pure AdS background and the metric fields play an important role. Another difference with \cite{FodorFG14} is the fact that the resonances are less numerous, occurring in higher order modes and narrower. All those reasons makes them relatively more difficult to find. We are very confident that this is a genuine effect but a more detailed study would be required to enhance our comprehension of those resonances.

\section{Time evolution}\label{s:evolution}

\subsection{Field equations}

In the following sections of the paper we use the same coordinate system and variables as in \cite{bizon-11,jalmuzna-11}. We look for spherically symmetric solutions with metric
\begin{equation}
\mathrm{d}s^2=\frac{L^2}{\cos^2 x}\left(-Ae^{-2\delta}\mathrm{d}\tilde{t}^{\;2}+\frac{1}{A}\mathrm{d}x^2
+\sin^2 x \, \mathrm{d}\Omega_{d-1}^2\right) , \label{sphgenm}
\end{equation}
where $L$ is related to the cosmological constant $\Lambda$ by \eqref{e:lambda}, and $A$ and $\delta$ are functions of the coordinates $\tilde{t}$ and $x$. We assume that the center of symmetry is regular at $x=0$ and that the metric approaches that of the anti-de Sitter one at infinity $x=\pi/2$. With the metric form choice \eqref{sphgenm} the anti-de Sitter metric is simply $A=1$ and $\delta=0$. The relation with the metric form \eqref{e:metdp1} follows from $\Psi^2 r=L\tan x$, and since in general $\Psi$ depends on both $t$ and $r$, the constant $r$ surfaces are different from the constant $x$ ones. Similarly, the constant $t$ and $\tilde{t}$ hypersurfaces are not the same either. However, at the moment of time reversal symmetry, the constant $t=0$ and $\tilde{t}=0$ surfaces agree, which makes easier the comparison of the results obtained with the two different coordinate system choices. We note however, that the calculation of Fourier components in time, such as those shown on Fig.~\ref{f:ori}, \ref{f:data_2.3} and \ref{f:reson}, is coordinate system dependent, and the comparison of Fourier modes in two different systems can be particularly difficult.

By Birkhoff's theorem, the only spherically symmetric vacuum solution is the Schwarzschild anti-de Sitter metric, which is given by $\delta=0$ and
\begin{equation}
A=1-m_S\frac{\cos^d x}{\sin^{d-2}x} \ , \label{asch}
\end{equation}
where the constant $m_S$  is related to the total mass $M_S$ of the Schwarzschild anti-de Sitter spacetime by $M_S=\gamma m_S$, with
\begin{equation}
\gamma=\frac{d-1}{16\pi G}\Omega_{d-1}L^{d-2} \ ,
\end{equation}
and the surface area of the $d-1$ dimensional sphere is $\Omega_{d-1}=2\pi^{d/2}/\Gamma(d/2)$.

In general spherically symmetric spacetimes there is a naturally defined radius function $R$, defined in terms of the area of the symmetry spheres. Using this function one can construct the Misner-Sharp energy (or local mass) function, which for negative cosmological constant can be defined as
\begin{equation}
M=\gamma m \ ,
\end{equation}
where
\begin{equation}
m=\frac{r^{d-2}}{L^{d-2}}
\left(1-g^{\mu\nu}r_{,\mu}r_{,\nu}+\frac{r^2}{L^2}\right) . \label{genmass}
\end{equation}
For the Schwarzschild anti-de Sitter metric $M=M_S$, constant. At infinity the mass function $M$ tends to the Abbott-Deser mass $M_{AD}=\gamma m_{AD}$ \cite{abbott}. In the coordinate system \eqref{sphgenm} the radius function is $R=L\tan x$, and the function $m$ can be expressed by the metric function $A$ as
\begin{equation}
m=\frac{\sin^{d-2}x}{\cos^d x}(1-A) \ . \label{eqmfunct}
\end{equation}
This shows that in order to have a finite total mass, the function $A$ must tend to $1$ at infinity $x=\pi/2$ as $(\pi/2-x)^d$.

Introducing the variables
\begin{equation}
 \Phi=\frac{\partial\phi}{\partial x} \quad , \qquad
 \Pi=\frac{e^\delta}{A}\frac{\partial\phi}{\partial \tilde{t}} \ ,
\end{equation}
the wave equation \eqref{waveeqgen} in the coordinate system \eqref{sphgenm} can be written into the form
\begin{equation}
\frac{\partial\Pi}{\partial \tilde{t}}
=\frac{1}{\tan^{d-1}x}\,\frac{\partial}{\partial x}\left(
\frac{A\tan^{d-1}x}{e^\delta}\Phi\right) . \label{waveeq}
\end{equation}
The equality of mixed partial derivatives of $\phi$ yields the evolution equation for $\Phi$,
\begin{equation}
 \frac{\partial\Phi}{\partial \tilde{t}}=\frac{\partial}{\partial x}
 \left(\frac{A}{e^\delta}\Pi\right) . \label{mixpardeq}
\end{equation}

The $(\tilde{t},\tilde{t})$ and the $(x,x)$ components of the Einstein's equations give
\begin{align}
\frac{\partial\delta}{\partial x}&=-\frac{8\pi G}{d-1}
\sin x\cos x\left(\Phi^2+\Pi^2\right) , \label{dxeq}\\
\frac{\partial A}{\partial x}&=A\frac{\partial\delta}{\partial x}
+\left[d\tan x+(d-2)\cot x\right](1-A) \ . \label{axeq}
\end{align}
If $\Phi$ and $\Pi$ are given on a constant $\tilde{t}$ hypersurface, then the metric variable $\delta$ can be calculated by integrating \eqref{dxeq}, and $A$ can be obtained by solving the differential equation \eqref{axeq}. Equation \eqref{dxeq} shows that it is advantageous to use units (or rescale $\phi$) such that
\begin{equation}
8\pi G=d-1 \ ,
\end{equation}
which we will assume in the next sections, and by this choice we will agree with the notation of \cite{jalmuzna-11}. The $(\tilde{t},x)$ component of the Einstein's equations give
\begin{equation}
\frac{\partial A}{\partial \tilde{t}}=-\frac{16\pi G}{d-1}\sin x\cos x\frac{A^2}{e^{\delta}}\Phi\Pi
 \ , \label{ateq}
\end{equation}
which is not independent from the previous equations. The remaining angular components of the Einstein's equations turn out to be equivalent to the wave equation \eqref{waveeq} after substituting the derivatives of $A$ and $\delta$ from \eqref{dxeq}, \eqref{axeq} and \eqref{ateq}.

The spatial derivative of $A$, as given by the right hand side of \eqref{axeq}, is not independent of $A$, but if we define $\mu=m e^{-\delta}$, i.e.
\begin{equation}
\mu=\frac{\sin^{d-2}x}{\cos^{d}x}\,\frac{1-A}{e^\delta} \ , \label{mudef}
\end{equation}
then the expression for its derivative becomes $\mu$ independent\cite{maliborski1},
\begin{equation}
 \frac{\partial\mu}{\partial x}=\frac{8\pi G}{d-1}\,
 \frac{\tan^{d-1}x}{e^\delta}\left(\Phi^2+\Pi^2\right) . \label{eqdmudx}
\end{equation}
If $\delta$ is already calculated by \eqref{dxeq}, then $\mu$ can be obtained by integrating this equation, and next, $A$ can be calculated by a simple algebraic operation using \eqref{mudef}.

Let us denote the coordinate distance from infinity by $y=\pi/2-x$. It is easy to check, that for a fixed anti-de Sitter background, i.e.~for $A=1$ and $\delta=0$, the leading order behavior of a massless scalar field $\phi$ is either $y^0$ or $y^d$. It can be shown that if a finite mass can be associated with the self gravitating system, i.e.~$A-1$ tends to zero at $y=0$ as $y^d$, as in \eqref{asch}, then the mass is time independent, and $\phi$ must necessarily tend to a time independent constant at infinity, which can be set to zero. The expansion of the functions then starts as
\begin{align}
\phi&=\phi_d y^d+\mathcal{O}(y^{d+2}) \ , \\
A&=1-m_{AD} y^d+\mathcal{O}(y^{d+2}) \ , \label{aexpans}\\
\delta&=\delta_\infty+\mathcal{O}(y^{2d}) \ ,
\end{align}
where the constant $m_{AD}$ is the Abbott-Deser mass of the system, while $\phi_d$ and $\delta_\infty$ are functions of $\tilde{t}$. All higher coefficients of the powers of $y$ in $\phi$, $A$ and $\delta$ can be expressed in terms of $m_{AD}$, $\phi_d$, $\delta_\infty$ and their derivatives. For even $d$ only even powers of $y$ occur in the expansion of $\phi$, $A$ and $\delta$, but for odd $d$ both odd and even powers occur. Introducing a new time coordinate as some function of $\tilde{t}$ it is always possible to set
\begin{equation}
\delta_\infty=0 \ ,
\end{equation}
which we assume in the following, as it was also done in \cite{craps,buchel-12,buchel-13}. This means that asymptotically $\tilde{t}$ will give the standard anti-de Sitter time coordinate. Alternatively, it would be possible to set $\delta=0$ at the center $x=0$, as it was done in \cite{bizon-11,craps14,okawa15}. In that case $L \tilde{t}$ would give the proper time at the center, but the $g_{\tilde{t}\tilde{t}}$ component of the metric would be time dependent at large distances from our localized object. In the following we will always assume that $\delta$ tends to zero at infinity.

We look for solutions which are regular at the center of symmetry $x=0$, which requires that $A$ tends to $1$ there, moreover $\delta$ and $\phi$ tend to some time dependent finite values. It also follows that in the power series expansions of $A$, $\delta$ and $\phi$ only even powers of $x$ appear, which implies that the functions have mirror symmetry there.

\subsection{Time-evolution code}

Our numerical code, which we use for the calculation of the time evolution of spherically symmetric configurations, is a fourth order method of lines code very similar to the one employed in \cite{bizon-11,jalmuzna-11,maliborski1}, which is described in detail in \cite{maliborski2}. The coordinate system \eqref{sphgenm} is used, and the funtions $A$, $\delta$, $\Phi$ and $\Pi$ depend on the coordinates $\tilde{t}$ and $x$.

We use a uniform grid with grid spacing $\Delta x$, and whenever possible we use symmetric fourth-order stencils to calculate radial derivatives. This can always be done near the center by taking into account that the functions $\phi$, $A$ and $\delta$ are symmetric at $x=0$. Near the outer boundary at $x=\pi/2$ we use the known asymptotic behaviour for the calculation of derivatives. If a function goes to zero as $f=y^\alpha(c_0+c_2 y^2+\cdots)$, where $\alpha$ is a known positive integer, then we calculate $c_0$ and $c_2$ from the value of $f$ at the last two grid points before the boundary, and for the derivative we substitute the corresponding value of $-f'$.

If we need to calculate the radial derivative of a product, such as for example that of $Ae^{-\delta}\Pi$ in the evolution equation \eqref{mixpardeq} for $\Phi$, then instead of expanding by the product rule of derivatives and calculating separately the derivative of $\Pi$, $A$ and $\delta$, we calculate first the value of the product $Ae^{-\delta}\Pi$ at each grid point and calculate the derivative from that. This way it is possible to obtain a more stable numerical method. The equation \eqref{waveeq} can be similarly used in order to obtain the time-evolution of $\Pi$, but it generates numerical instabilities both at the center and at the outer boundary. Hence it is used only for $\pi/8\leq x\leq 3\pi/8$. For $x<\pi/8$ and for $x>3\pi/8$ we apply the method described in \cite{maliborski2} by rewriting \eqref{waveeq} into the form
\begin{equation}
\frac{\partial\Pi}{\partial \tilde{t}}
=\left(1+\frac{d-1}{\cos(2x)}\right)
\frac{\partial}{\partial x}\left(\frac{A\Phi}{e^\delta}\right)
-\frac{d-1}{2}\tan(2x)\frac{\partial}{\partial x}
\left(\frac{A\Phi}{e^\delta\sin x\cos x}\right) . \label{eqpit2}
\end{equation}
This way we only have to calculate the derivatives of functions which are less singular at the boundaries, and we obtain a stable numerical method. In the last term of \eqref{eqpit2} the value of $\Phi/\sin x$ at the center $x=0$ is substituted by the derivative of $\Phi$ using l'H\^opital's rule. Except for the last two grid points before the boundary at $x=\pi/2$, dissipative terms proportional to the sixth radial derivatives of $\Phi$ and $\Pi$ are added to the time-evolution equations, in order to make sure that the numerical evolution is stable. By choosing both proportionality factors as $\sigma(\Delta x)^5$, the numerical method remains fourth order, and the value of the constant $\sigma$ can be chosen, for example, to be $0.03$.

Time-evolution of $\Phi$ and $\Pi$ are calculated at each grid point by a fourth-order Runge-Kutta scheme. At each time step, including the Runge-Kutta substeps, $\delta$ is calculated by numerically integrating the right-hand side of \eqref{dxeq} from the outer boundary, where $\delta=0$. The variable $A$ is calculated by first integrating \eqref{eqdmudx} from the center, where $\mu=0$, and then calculating $A$ using \eqref{mudef}. This way we avoid integrating the differential equation \eqref{axeq} for $A$ by a Runge-Kutta scheme, which would require getting midpoint values at $\Delta x/2$ by numerical interpolation. In our method, the forth order accurate numerical integration of $\delta$ and $\mu$ is done by fitting third order polynomials at four neighboring grid points, and using the integral of this interpolating function between two neighboring grid points. The asymptotic behavior of the integrand of $\delta$ at the outer boundary is $f=y^{2d-1}(c_0+c_2 y^2)$, while the behavior of $\mu$ near the center is of the form $f=x^{d-1}(c_0+c_2 x^2)$. The constants $c_0$ and $c_1$ are calculated from the values of the integrand at the first two points, and the integral up to these points is represented by the integral of $f$. The typical number of spatial grid points used on our numerical simulations were between $2^{8}$ and $2^{14}$ points.

\subsection{Results obtained by the time-evolution code}

We have used several configurations obtained by the spectral numerical code in Sec.~\ref{sec:periodic} as initial data for our time-evolution code. We take the field value at the time slice where the configuration is time reversal symmetric, i.e when $\Pi=0$. Only the field $\Phi$ is taken from the spectral code, appropriately changing the radial coordinates from $r$ to $x$. Values of $A$ and $\delta$ are calculated by the same method on the initial time-slice as on all the later time-slices. Firstly, this procedure confirms that the time evolution of the initial data indeed leads to time-periodic solutions with the expected oscillation frequency, precise to several digits. Secondly, it enables us to study the stability properties of these configurations. 

\begin{figure}[!hbtp]
\includegraphics[width=8cm,clip]{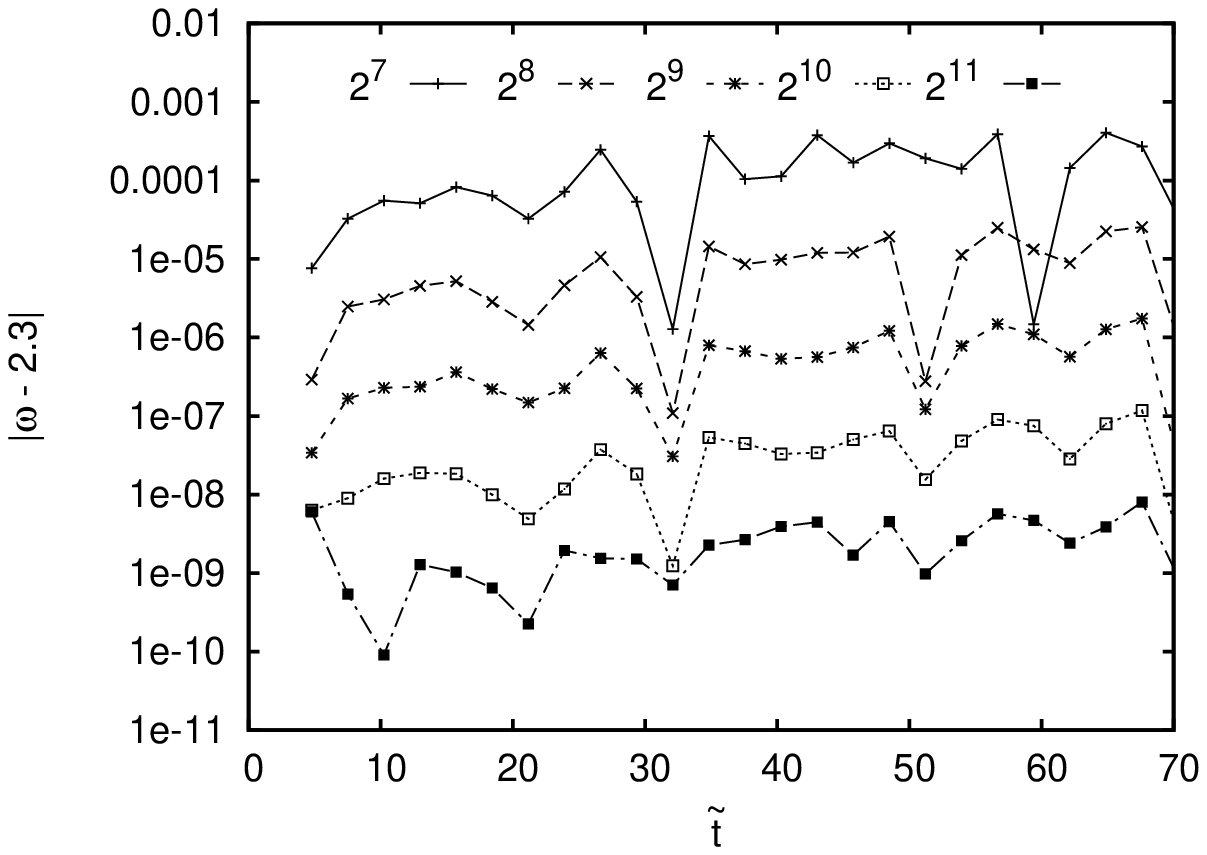}
\includegraphics[width=8cm,clip]{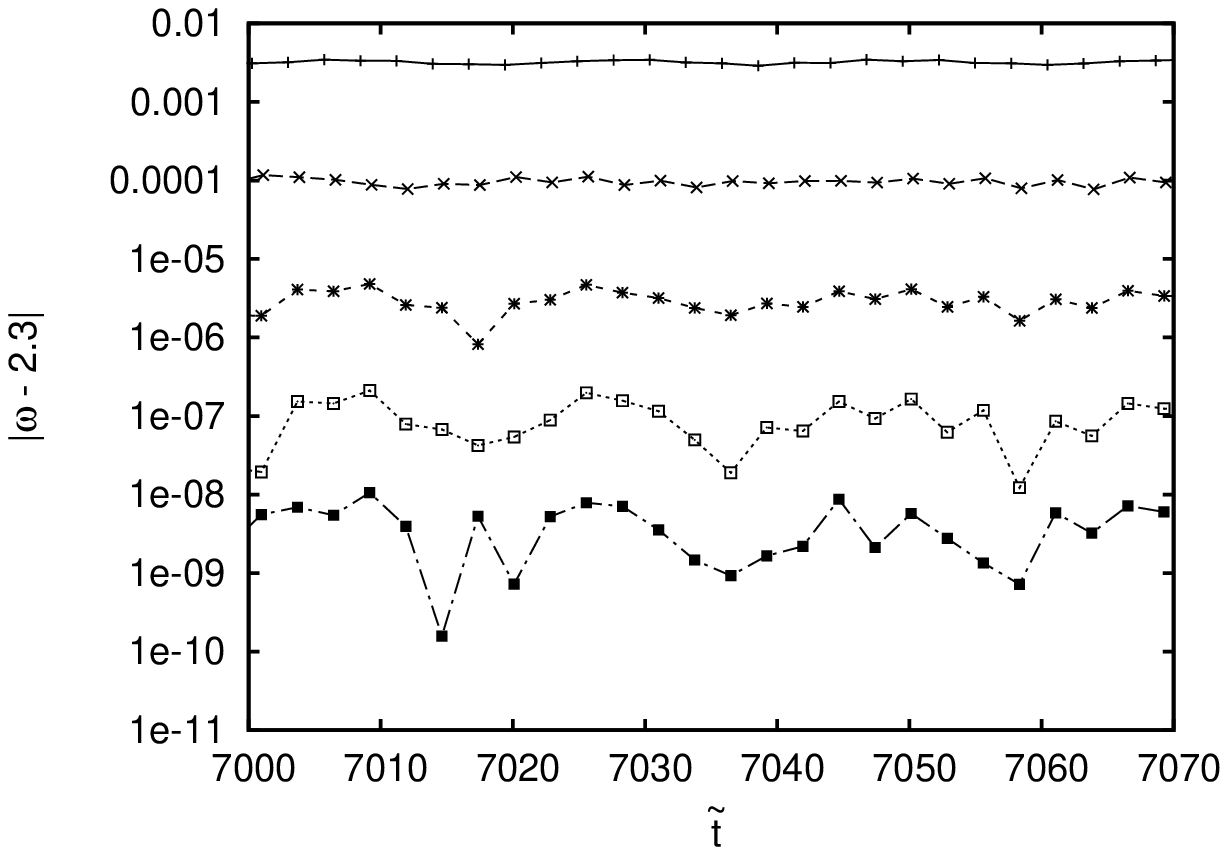}
\caption{\label{f:freqerr} Difference from the expected frequency of the oscillations as a function of time, evolving initial data corresponding to $\omega=2.3$ and $d=3$. An initial and a much later period is shown for numerical resolutions from $2^7$ to $2^{11}$ points .}
\end{figure}

Small amplitude configurations turn out to be always stable. These have frequency $\omega$, which is slightly lower than the frequency $\omega_0=d$ of the nodeless linearized solution. These remain oscillating with essentially constant frequency in the time evolution code for arbitrary long time-periods. For higher amplitude configurations, the oscillation frequency is lower, but the behavior remains very similar. In order to illustrate this, on Fig.~\ref{f:freqerr} we show the time evolution of the frequency when numerically evolving initial data corresponding to $\omega=2.3$, in case of $d=3$, with increasing resolutions in the time-evolution code.
The frequency is obtained by calculating the time difference between the current and the previous maximums of the oscillating scalar field at the center.

\begin{figure}[!hbtp]
\includegraphics[width=8cm,clip]{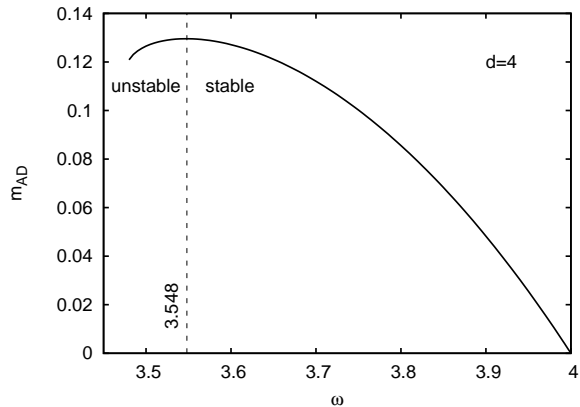}
\caption{\label{f:mass4d} Frequency dependence of the mass of the configurations for $d=4$ spatial dimensions.  }
\end{figure}

However, as the amplitude increases and the oscillation frequency becomes lower, below a certain frequency $\omega_s$, the configurations become unstable. For $d=3$ spatial dimensions $\omega_s=2.253$, and for $d=4$ the instability starts below $\omega_s=3.548$. In both cases this corresponds to the frequency where the mass of the configuration is maximal, as it is usually the case for physically relevant localized objects. We note however, that for self interacting scalar fields on a fixed AdS background instability generally occurs already before the energy of the configuration reaches a maximum\cite{FodorFG14}. On Fig.~\ref{f:mass4d} we show the frequency dependence of the mass $m_{AD}$ in the $d=4$ dimensional case. The mass is calculated from the results of the spectral code using \eqref{genmass} at some large radius.
A similar figure for $d=3$ will be given in Sec.~\ref{sechighord} together with the corresponding small-amplitude expansion results.

In the $d=4$ case $\omega_s=2.253$ corresponds to the frequency $\Omega_s=6.325$ as measured by a central observer (see Fig.~\ref{f:omom}). The configurations represented by the circles on Fig.~\ref{f:comppolish}, which were taken from \cite{maliborski1}, are all in the stable domain, instability only occurs for higher amplitude configurations. We note that the instability does not appear at the turning point where $\varepsilon$ is maximal on Fig.~\ref{f:comppolish}, it begins at somewhat higher $\Omega$.

The narrow resonances that we have found using the spectral numerical code in Sec.~\ref{ss:results} are all in the unstable domain, both for $d=3$ and $d=4$.

The unstable solutions initially oscillate with the expected frequency, however after about a few dozen oscillations the oscillation amplitude starts changing, and eventually the scalar field collapses into a black hole. The metric function $A$ can be seen to approach zero at some radius $x$, which signals the appearance of an event horizon. The more precisely the initial data is given, the longer the configuration oscillates, and the later the collapse occurs, which indicates that very likely there is a perturbation mode which makes these high-amplitude periodic solutions unstable.

\begin{figure}[!hbtp]
\includegraphics[width=8cm,clip]{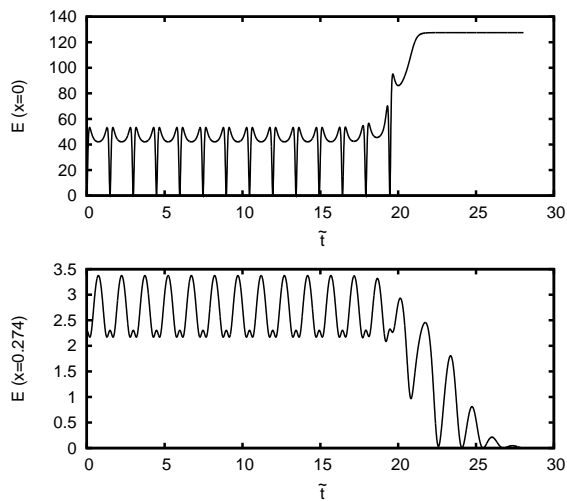}
\caption{\label{f:endens} Time evolution of the energy density $E$ for an initial data  which is in the large amplitude unstable domain, corresponding to $\omega=2.104$. The upper panel shows $E$ at the center, while the lower one is at a radius that is larger than the one where the horizon forms. }
\end{figure}

As a concrete example, on Fig.~\ref{f:endens} we show the time evolution of the energy density
\begin{equation}
E=\frac{A\cos^2x}{2L^2}\left(\Phi^2+\Pi^2\right)
\end{equation}
at two places for initial data belonging to the frequency $\omega=2.104$. The upper plot shows $E$ at the symmetry center, while the lower one at a radius which is outside the sphere where the event horizon first appears.
For about six oscillations the evolution appears to be periodic, but then the unstable mode grows large enough to destroy the periodicity, and the scalar field quickly collapses into a black hole. The lower panel shows that the energy density tends to zero outside the horizon, while the upper panel shows that it approaches a constant value at the center. However, this does not mean that the energy density actually stops increasing at the center, it is only the time foliation which slows down very much, because the metric function $\delta$ increases to large values in the central region. Our coordinate system, which is based on a constant area-radius foliation, breaks down when an apparent horizon forms. The specific way of how a configuration collapses to a black hole, and after how many oscillations, depends strongly on what kind of small perturbations are already present in the initial data, and also on how strong is the instability for the given frequency. The large amplitude configuration which we show on Fig.~\ref{f:endens} is relatively short living even if the initial data is calculated with very high precision by the spectral code.

\section{Small-amplitude expansion} \label{sec:smallampl}

\subsection{Linearized equations}

We expand around the vacuum anti-de Sitter metric using the coordinate system \eqref{sphgenm}. We use our previously developed perturbative framework for oscillons \cite{FFHM,moredim} and oscillatons \cite{oscillaton10,oscillatons}. In terms of a small parameter $\varepsilon$ we write
\begin{equation}
\phi=\sum_{n=1}^\infty\phi^{(n)}\varepsilon^n \ , \quad
A=1+\sum_{n=1}^\infty A^{(n)}\varepsilon^n \ , \quad
\delta=\sum_{n=1}^\infty\delta^{(n)}\varepsilon^n \ , \label{epsexpansion}
\end{equation}
where $\phi^{(n)}$, $A^{(n)}$ and $\delta^{(n)}$ are functions of $\tilde{t}$ and $x$. We substitute this expansion into the wave equation \eqref{waveeq}, and into the Einstein's equations components \eqref{dxeq} and \eqref{axeq}. Since in many cases it is easier to solve, we also use \eqref{ateq}, but we cannot eliminate the use of \eqref{axeq} completely. It will turn out later, that the expansion of $\phi$ contains only odd powers of $\varepsilon$, while the expansion of $A$ and $\delta$ only even powers of $\varepsilon$. Assuming that there is only a single one-parameter family of solutions that we consider, this follows from the $\phi\rightarrow -\phi$ symmetry of the system. We also note that our expansion parameter $\varepsilon$ is not the same as the parameter $\varepsilon$ used in \cite{maliborski1}, although to leading order they are proportional to each other.

Since the Einstein's equations contain $\phi$ quadratically, to first order in $\varepsilon$ the equations for $A$ and $\delta$ are the same as in vacuum. Since the only spherically symmetric vacuum solution is the Schwarzschild anti-de Sitter metric, which only has a regular center if the mass is zero, it follows that $A^{(1)}=0$ and $\delta^{(1)}=0$. From the wave equation \eqref{waveeq} it follows that the leading order coefficient of the scalar field satisfies the differential equation
\begin{equation}
\frac{\partial^2\phi^{(1)}}{\partial \tilde{t}^2}=\frac{\partial^2\phi^{(1)}}{\partial x^2}
+\frac{d-1}{\sin x\cos x}\frac{\partial\phi^{(1)}}{\partial x} \ .
\end{equation}
Let us look for periodically oscillating solutions in the form $\phi^{(1)}=p\cos(\omega \tilde{t})$, where $p$ only depends on $x$. Since we assume that $\delta$ tends to zero at infinity, the time coordinate $\tilde{t}$ is the standard anti-de Sitter time, and the frequency $\omega$ represents the frequency measured by some faraway static observer. However, since the proper time at the center of the vacuum anti-de Sitter spacetime in the coordinate system \eqref{sphgenm} is not $\tilde{t}$ but $L \tilde{t}$, the physical frequency of the oscillating solutions is actually not $\omega$ but $\omega/L$. Then $p$ has to satisfy the ordinary differential equation
\begin{equation}
\frac{\mathrm{d}^2p}{\mathrm{d}x^2}
+\frac{d-1}{\sin x\cos x}\,\frac{\mathrm{d}p}{\mathrm{d}x}
+\omega^2 p=0 \ . \label{linpeq}
\end{equation}
Solutions, which are regular at the center and localized in the sense that $p$ tends to zero at infinity, only exist for certain discrete values of the frequency, $\omega=\omega_n$, where
\begin{equation}
\omega_n=d+2n \ , \label{omegaupn}
\end{equation}
for $n\geq0$ integers. The corresponding regular localized solutions can be given in the form \cite{avisisham}
\begin{equation}
p_n=\frac{n!}{(d/2)_n}\cos^d x\,
P_{n}^{(d/2-1,d/2)}\left(\cos(2x)\right) ,  \label{pneqgen}
\end{equation}
where $P$ denotes the Jacobi polynomial, and $(\alpha)_n=\alpha(\alpha+1)\ldots(\alpha+n-1)$ denotes the Pochhammer symbol, with $(\alpha)_0=1$. The solutions $p_n$ are normalized in order to make their central values $1$  at $x=0$. The integer $n$ labels the number of nodes of the solutions. The first few functions are
\begin{align}
p_0&=\cos^d x \ , \\
p_1&=\frac{\cos^d x}{d}\left[(d+1)\cos(2x)-1\right] , \\
p_2&=\frac{\cos^d x}{2d}\left[(d+3)\cos(4x)-4\cos(2x)+d+1\right] , \\
p_3&=\frac{\cos^d x}{4d(d+2)}\left\{(d+3)\left[(d+5)\cos(6x)
-6\cos(4x)+3(d+1)\cos(2x)\right]-6(d+1)\right\} .
\end{align}
We note that for odd $d$ it may be possible to write \eqref{pneqgen} in terms of trigonometric functions without using orthogonal polynomials. For example, for $d=3$
\begin{equation}
p_n=\frac{1}{4\sin x}\left(\frac{\sin[2(1+n)x]}{1+n}+
\frac{\sin[2(2+n)x]}{2+n}\right) .
\end{equation}
Rescaling the polynomials as
\begin{equation}
e_n=\frac{2\sqrt{(n+d-1)!}}{\sqrt{n!}\,\Gamma\left(\frac{d}{2}\right)}p_n \ ,
\end{equation}
the functions $e_n$ are normalized in order to make them an orthonormal base on the Hilbert space $L^2([0,\pi/2],\tan^{d-1}\mathrm{d}x)$.

To the linear approximation, the general localized regular solution can be written as a sum of these solutions with arbitrary amplitudes $a_n$ and phases $b_n$,
\begin{equation}
\phi=\sum_{n=0}^\infty a_n\cos(\omega_n \tilde{t}+b_n)p_n \ .
\end{equation}
For the linearized problem all these solutions give stable configurations, and since all frequencies are integers, the whole solution is periodic with frequency $\omega=1$ or $2$. To the linear approximation, the frequency of each mode is independent of its amplitude. However, nonlinear effects will make the frequencies amplitude dependent. In the small-amplitude expansion procedure this means that each $\omega_n$ becomes $\varepsilon$ dependent. In general, one expects that each $\omega_n$ will change independently, becoming unrelated real numbers, and therefore the solution will stop being periodic. In Refs.~\cite{bizon-11} and \cite{buchel-12} it was argued that proceeding to higher orders in the $\varepsilon$ expansion secularly growing terms, of the type $\tilde{t}\cos(\omega \tilde{t})$, appear in all cases when two or more, but only a finite number of $a_n$ are nonzero. According to \cite{buchel-12} in the case when an infinite number of $a_n$ are different from zero, it may be possible to cancel all secular terms. In that case however, the solution will be quasiperiodic, consisting of infinite number of modes with not rationally related frequencies \cite{balasubramanian14}.

Since periodic solutions are of a great interest in their own, in this paper we concentrate on constructing those. If only a single $a_n$ is nonzero, then all secular terms can be absorbed by changing $\omega_n$ as a function of $\varepsilon$. This means that there is a one-parameter family of nonlinear solutions emerging from each $p_n$ mode, and their frequency gradually becomes more and more different from $\omega_n$ as the amplitude grows. We will be mainly interested in the family of solutions starting from the nodeless mode $p_0$, since that one is expected to be the most stable, and also the one with the lowest mass.

\subsection{Periodic solutions}

In this paper we concentrate on periodic solutions, having a definite frequency $\omega$, which generally depends on the oscillation amplitude. In order to include this periodicity in our formalism more conveniently, we introduce a new time coordinate
\begin{equation}
\tau=\omega \tilde{t} \ .
\end{equation}
In terms of $\tau$ the oscillation frequencies of all Fourier modes will be integers. Using $\tau$ each time derivative in the field equations \eqref{waveeq}-\eqref{ateq} will bring an $\omega$ factor, and there will be $\omega^2$ terms in the equations. We take into account the $\varepsilon$ dependence of $\omega$ by expanding it as
\begin{equation}
\omega=\omega^{(0)}\left(1+\sum_{j=1}^\infty\omega^{(j)}\varepsilon^j\right) \ ,
\end{equation}
where $\omega^{(0)}$ and $\omega^{(j)}$ are some numbers. It will turn out later, that $\omega^{(j)}=0$ for odd $j$. Assuming that there is only a one-parameter family of solutions that we consider, this also follows from the $\phi\rightarrow -\phi$ symmetry of the system.

In the zero amplitude limit $\omega$ should tend to one of the $\omega_n$ frequencies, hence $\omega^{(0)}=\omega_n$ for some $n$. From now on we concentrate on the solution emerging from the nodeless mode, so we assume that
\begin{equation}
 \omega^{(0)}=\omega_0=d \ .
\end{equation}

We Fourier decompose the coefficients in the small-amplitude expansion of $\phi$, $A$ and $\delta$ as
\begin{equation}
\phi^{(n)}=\sum_{j=0}^\infty\phi^{(n)}_j\cos(j\tau) \ , \quad
A^{(n)}=\sum_{j=0}^\infty A^{(n)}_j\cos(j\tau) \ , \quad
\delta^{(n)}=\sum_{j=0}^\infty\delta^{(n)}_j\cos(j\tau) \ ,
\end{equation}
where $\phi^{(n)}_j$, $A^{(n)}_j$ and $\delta^{(n)}_j$ are functions depending on $x$. We could also have included $\sin(j\tau)$ terms in the expansion, but because of the time reversal symmetry of the problem, those terms would turn out to be zero anyway. It will also turn out later that $\phi^{(n)}_j=0$ for even $j$, while $A^{(n)}_j=0$ and $\delta^{(n)}_j=0$ for odd $j$. This is closely related to the $\phi\rightarrow -\phi$ symmetry.

Substituting the expansion into the wave equation \eqref{waveeq}, the equations for all $\phi^{(n)}_j$ take the form
\begin{equation}
\frac{\mathrm{d}^2\phi^{(n)}_j}{\mathrm{d}x^2}
+\frac{d-1}{\sin x\cos x}\,\frac{\mathrm{d}\phi^{(n)}_j}{\mathrm{d}x}
+j^2 d^2 \phi^{(n)}_j=F^{(n)}_j \ , \label{phijneq}
\end{equation}
where $F^{(n)}_j$ depends only on the lower $\varepsilon$ order functions, $\phi^{(l)}_k$, $A^{(l)}_k$, $\delta^{(l)}_k$ with $l<n$. For the first Fourier modes, i.e.~for $j=1$, the homogeneous part of the equation is always solved by $\phi^{(n)}_1=\cos^d x$, consequently one can always add a constant times $\cos^d x$ to a given inhomogeneous solution. We use this freedom to specify how we parametrize the various states by the parameter $\varepsilon$. We set
\begin{align}
\phi^{(1)}&=1 \text{ at } x=0 \ , \ \tilde{t}=0 \ , \label{phi1x0} \\
\phi^{(n)}&=0 \text{ at } x=0 \ , \ \tilde{t}=0 \text{ for } n\geq2 \ . \label{phinx0}
\end{align}
By this choice the parameter $\varepsilon$ gives the central value of the scalar field $\phi$ at the time when the configuration is time reversal symmetric, i.e.~at $\tilde{t}=0$. We note that our parameter $\varepsilon$ is different from the parameter $\varepsilon$ used in \cite{maliborski1}, which corresponds to the scalar product of $\phi$ at $\tilde{t}=0$ with the nodeless linear solution $e_0$. Our $\varepsilon$ is a monotonous function of the frequency in all the domain where we could calculate periodic states with the spectral numerical code, both in $d=3$ and $d=4$ dimensions.

\subsection{First order in the small-amplitude expansion}

To the leading order in the small-amplitude expansion, i.e.~for $n=1$, obviously all $F^{(1)}_j$ in \eqref{phijneq} are zero. Comparing to \eqref{linpeq}, we can see that the regular localized solution for $\phi^{(1)}_1$ is an arbitrary constant times $p_0$. However, other $\phi^{(1)}_j$ may also have localized regular solutions after setting the basic frequency $\omega^{(0)}=d$. According to \eqref{omegaupn}, this happens if the frequency $jd$ is equal to $d+2n$ for some $n\geq0$ integer, i.e.~if $(j-1)d=2n$. If the number of spatial dimensions $d$ is even, then this holds for all $j\geq 1$, and for odd $d$ there are solutions when $j$ is an odd number. If they exist, the regular localized solutions for $\phi^{(1)}_j$ are given by multiples of $p_n$, and have $(j-1)d/2$ nodes. The sum of all these solutions, with arbitrary phase shifts and amplitudes, are valid solutions of the linearized problem. However, continuing to higher order in the $\varepsilon$ expansion, assuming that $\phi^{(1)}_1$ is nonzero, it turns out that at $\varepsilon^3$ order there are localized regular solutions only if
\begin{equation}
\phi^{(1)}_j=0 \quad \mathrm{if} \quad j\neq 1 \ . \label{phi1jeq}
\end{equation}
We have checked this for $d=3$, but we expect that it holds for any dimensions. For example, if we allow that $\phi^{(1)}=c_1 p_1\cos\tau+c_3 p_3\cos(3\tau)$ for some constants $c_1$ and $c_3$ in the $d=3$ case, then at $\varepsilon^3$ order it turns out that for $\phi^{(3)}_5$ there are localized regular solutions only if $c_1c_3^2=0$.
The third order secular terms have been explicitly calculated in Ref.~\cite{craps14}, where it was noted that most of the potentially resonant terms do not in fact arise. In the language of \cite{craps14}, the solution is quasiperiodic if the renormalized amplitudes remain constant. Periodicity with a single frequency requires further restrictions on the renormalized phases and the resulting frequencies. It is a promising strategy to find a proof, using the methods of Ref.~\cite{craps14}, to show that exact periodicity is possible if and only if there is one single mode present in the linear order of the expansion, for arbitrary dimensions. This problem is beyond the scope of the present paper and is left for future research.
From now on we assume that \eqref{phi1jeq} holds. Since we plan to interpret $\varepsilon$ as the central amplitude of $\phi$ at $\tilde{t}=0$, according to \eqref{phi1x0} we take
\begin{equation}
\phi^{(1)}_1=\cos^d x \ .
\end{equation}

Since to first order the metric components and the scalar field decouple, the only regular solution is the anti-de Sitter metric, so necessarily
\begin{equation}
A^{(1)}_j=0 \quad , \quad \delta^{(1)}_j=0 \ .
\end{equation}
Proceeding to higher orders in the small-amplitude expansion, it will turn out similarly, that
\begin{equation}
A^{(n)}_j=0 \quad , \quad \delta^{(n)}_j=0 \quad \text{for odd } n \ .
\end{equation}

\subsection{Second order in the small-amplitude expansion: metric variables}

Equation \eqref{dxeq} gives the derivatives of $\delta^{(2)}_j$ in terms of lower order quantities, which can generally be integrated easily. Since we would like $\tilde{t}$ to give the anti-de Sitter coordinate time for faraway static observers, we set the integration constants in order to make $\delta^{(n)}_j=0$ at $x=\pi/2$. The only nonvanishing components at $\varepsilon^2$ order are
\begin{align}
\delta^{(2)}_0&=\frac{d}{4}\cos^{2d}x \ , \\
\delta^{(2)}_2&=\frac{d}{4(d+1)}\cos^{2d}x\left[1-d\cos(2x)\right] .
\end{align}
We remind the reader that we use the choice $8\pi G=d-1$ in this section of the paper.

Equation \eqref{ateq} gives algebraic equations for all $A^{(2)}_j$, except that it does not give any relation for $A^{(2)}_0$. The only nonvanishing component for $j>0$ is
\begin{equation}
A^{(2)}_2=\frac{d}{2}\sin^2 x\cos^{2d}x \ . \label{a22eq}
\end{equation}

For even $d$, the functions $\delta^{(2)}_0$, $\delta^{(2)}_2$ and $A^{(2)}_2$ can be written as sums of finite number of $p_n$ basis functions. However, for odd $d$ infinite number of $p_n$ is needed, and the convergence is very slow. The reason for this can be seen by power series expanding at infinity $x=\pi/2$. The expansion of each $p_n$ contains only $(x-\pi/2)^{d+2k}$ terms, where $k\geq 0$ integer. For odd $d$ these are all odd powers. However, the expansion of $\delta^{(2)}_0$, $\delta^{(2)}_2$ and $A^{(2)}_2$ contain $(x-\pi/2)^{2d+2k}$ terms which are all even powers, even for odd $d$.

After setting the components of $A^{(2)}_2$ according to \eqref{a22eq} and setting $A^{(2)}_j=0$ for all other $j>0$, the field equation \eqref{axeq} only yields one first order linear differential equation,
\begin{equation}
\frac{\mathrm{d}A^{(2)}_0}{\mathrm{d}x}+\frac{d-2\cos^2x}{\sin{x}\cos{x}}A^{(2)}_0
+\frac{d^2}{2}\sin x\cos^{2d-1}x=0 \ .
\end{equation}
This can be rewritten into the form
\begin{equation}
\frac{\mathrm{d}}{\mathrm{d}x}\left(\frac{\sin^{d-2}x}{\cos^d x}A^{(2)}_0\right)
+\frac{d^2}{2^d}\sin^{d-1}(2x)=0 \ , \label{a20eq2}
\end{equation}
so the problem is reduced to integrating integer powers of the sine function. For general $d$ the integral can be given in terms of hypergeometric functions, yielding
\begin{equation}
A^{(2)}_0=\frac{d^2\cos^d x}{2^{d+1}\sin^{d-2}x}\left[
\cos(2x)\,{}_2F_1\left(\frac{1}{2},1-\frac{d}{2},\frac{3}{2},\cos^2(2x)\right)
-\frac{\sqrt{\pi}\,\Gamma\left(\frac{d}{2}\right)}{2\Gamma\left(\frac{d+1}{2}\right)}
\right] .  \label{a20gen}
\end{equation}
The constant of integration was set in order to make $A^{(2)}_0$ regular at the center $x=0$. If the number of spatial dimensions $d$ is even, then the second argument of the hypergeometric function becomes a negative integer, and the hypergeometric series stops at finite order. In that case the hypergeometric function can be written in terms of a Gegenbauer polynomial. For even $d$
\begin{equation}
A^{(2)}_0=-\frac{d^2\sqrt{\pi}\,\Gamma\left(\frac{d}{2}\right)\cos^d x}
{2^{d+2}\Gamma\left(\frac{d+1}{2}\right)\sin^{d-2}x}\left[1+
C_{d-1}^{(1-d)/2}\left(\cos(2x)\right)\right] .
\end{equation}
The first few specific values for given even dimensions are
\begin{align}
d&=2: & A^{(2)}_0&=-\cos^2x\sin^2x \ , \\
d&=4: & A^{(2)}_0&=-\frac{2}{3}\cos^4x\sin^2x\left[2+\cos(2x)\right] \ ,\\
d&=6: & A^{(2)}_0&=-\frac{3}{40}\cos^6x\sin^2x\left[19+18\cos(2x)+3\cos(4x)\right] \ .
\end{align}
For odd $d$ it is possible to use identities to convert the hypergeometric function in \eqref{a20gen} into a sum of finite number of terms, but the result will not be a polynomial in $\sin x$ and $\cos x$. Instead of that approach, it is easier to apply directly the expression 2.513 1 in \cite{gradshsteyn} for the integral of even powers of the sine function, giving for odd $d$
\begin{equation}
A^{(2)}_0=-\frac{d^2\cos^d x}{2^{2d-1}\sin^{d-2}x}\left[
\binom{d-1}{\frac{d-1}{2}}x
+(-1)^{\frac{d-1}{2}}\sum_{k=0}^{\frac{d-3}{2}}(-1)^k\binom{d-1}{k}
\frac{\sin[2(d-1-2k)x]}{d-1-2k}
\right] .
\end{equation}
The first few specific values for odd spatial dimensions are
\begin{align}
d&=3: & A^{(2)}_0&=-\frac{9\cos^3x}{64\sin x}[4x-\sin(4x)] \ , \\
d&=5: & A^{(2)}_0&=-\frac{25\cos^5x}{2048\sin^3x}[24x-8\sin(4x)+\sin(8x)] \ ,\\
d&=7: & A^{(2)}_0&=-\frac{49\cos^7x}{49152\sin^5x}
[120x-45\sin(4x)+9\sin(8x)-\sin(12x)] \ .
\end{align}

For even $d$ the metric function $A^{(2)}_0$ can again be expressed as a finite sum of $p_n$ basis functions, but for odd $d$ infinite number of $p_n$ are required. In this case the power series expansion of $A^{(2)}_0$ for odd $d$ contains both odd and even powers of $x-\pi/2$ at infinity.

\subsection{Second order in the small-amplitude expansion: scalar field}

To $\varepsilon^2$ order the scalar field equation \eqref{waveeq} gives the same form of equations for $\phi^{(2)}_j$ as \eqref{phijneq}, and the only nonvanishing source term is
\begin{equation}
F^{(2)}_1=-2d^2\omega^{(1)}\cos^d x \ . \label{f21eq}
\end{equation}
From now on we take $\phi^{(n)}_j=0$ for all those $n$ and $j$ for which $F^{(n)}_j=0$. Even if there are regular localized solutions of the homogeneous problem, we take the trivial zero solution if it is possible. If we would not do this, we expect that there would be no localized and regular solutions for some quantities at two orders higher in the $\varepsilon$ expansion. We cannot give a general proof for this, but we have checked it for some concrete $n$, $j$ and $d$ values.

In order to calculate the solution of the inhomogeneous equation \eqref{phijneq}, we need two linearly independent solutions of the homogeneous problem. One homogeneous solution for $\phi^{(n)}_j$ is
\begin{equation}
u_1=\cos^d x\,{}_2F_1\left(\frac{d}{2}(1+j),\frac{d}{2}(1-j);1+\frac{d}{2};\cos^2x\right) ,
\end{equation}
which is always localized in the sense that it goes to zero at infinity as $(\pi/2-x)^d$. When $j\geq1$ and $d(j-1)$ is even, the hypergeometric series closes at finite order, and $u_1$ becomes proportional to one of the localized regular solutions defined in \eqref{pneqgen}
\begin{equation}
u_1=\frac{(-1)^n}{j}p_n \ , \quad n=\frac{d}{2}(j-1) \ .
\end{equation}
When $d$ is even, this happens for all $j\geq1$, while for odd $d$ there are localized solutions if and only if $j$ is odd. This is important, since it turns out that at each order in $\varepsilon$ there are nonzero $F^{(n)}_j$ source terms in \eqref{phijneq} only for odd $j$.

For odd $d$ a second solution can be given as
\begin{equation}
u_2={}_2F_1\left(\frac{dj}{2},-\frac{dj}{2};1-\frac{d}{2};\cos^2x\right)
=\frac{1}{\sin^{d-2}x}\ {}_2F_1\left(1-\frac{d}{2}(j+1),1+\frac{d}{2}(j-1);
1-\frac{d}{2};\cos^2x\right) . \label{u2dodd}
\end{equation}
This function is not defined when the third argument of the hypergeometric function, $1-d/2$, is a nonpositive integer. When both $d$ and $j$ are odd, the second form becomes a polynomial,
\begin{equation}
u_2=\frac{n!(-1)^n}{\left(1-\frac{d}{2}\right)_n\sin^{d-2}x}
P_n^{(1-d/2,-d/2)}\left(\cos(2x)\right) \ , \quad n=\frac{d}{2}(j+1)-1 \ ,
\end{equation}
where $P$ denotes the Jacobi polynomial and $()_n$ the Pochhammer symbol. For concrete dimensions it may be possible to find simpler expressions. For example for $d=3$ and for odd $j$
\begin{equation}
u_2=(-1)^{(j+1)/2}\left[3j\cos(3jx)\cot x+\sin(3jx)\right] .
\end{equation}
For $d=5$ and for odd $j$
\begin{equation}
u_2=\frac{(-1)^{(j+1)/2}}{3}\left[\frac{5j}{\sin^3x}\cos(3x)\cos(5jx)
+(25j^2\cot^2x-3)\sin(5jx)\right] .
\end{equation}

The general solution of the inhomogeneous differential equation \eqref{phijneq} can be given in terms of two independent solutions of the homogeneous problem, $u_1$ and $u_2$, as
\begin{equation}
\phi^{(n)}_j=u_2\int_{x_1}^x\frac{u_1}{W}F^{(n)}_j\mathrm{d}x
-u_1\int_{x_2}^x\frac{u_2}{W}F^{(n)}_j\mathrm{d}x \ ,  \label{phinjint}
\end{equation}
where $x_1$ and $x_2$ are some constants, and the Wronskian is
\begin{equation}
W=u_1\frac{\mathrm{d}u_2}{\mathrm{d}x}-u_2\frac{\mathrm{d}u_1}{\mathrm{d}x} \ .
\end{equation}
For odd $d$, when $u_2$ can be defined by \eqref{u2dodd}, the Wronskian can be calculated to be
\begin{equation}
W=d\cot^{d-1}x \ . \label{wgen}
\end{equation}
By Abel's differential equation identity, the Wronskian of any two solutions of the homogeneous equation is a constant times $\cot^{d-1}x$, and hence by appropriately scaling the second solution $u_2$ we can make \eqref{wgen} valid for even $d$ as well. For example, for $d=2$ we can choose the second solution as
\begin{align}
u_2^{(j=1)}&=1-2\cos^2x\log(\cot x) \ ,\\
u_2^{(j=3)}&=\frac{1}{4}\left[7+18\cos(2x)+15\cos(4x)\right]
-\frac{3}{2}\cos^2x\left[3-4\cos(2x)+5\cos(4x)\right]\log(\cot x) \ ,
\end{align}
and for $d=4$ we can choose
\begin{align}
u_2^{(j=1)}&=6+3\cos(2x)-\frac{2}{\sin^2x}-12\cos^4x\log(\cot x) \ ,\\
u_2^{(j=3)}&=\frac{1}{\sin^2x}-\frac{7}{128}\cot^2x
\left[104-186\cos(2x)+75\cos(6x)-360\cos(8x)+495\cos(10x)\right] \\
&\ \ -\frac{105}{16}\cos^4x\left[35-40\cos(2x)+60\cos(4x)-24\cos(6x)
+33\cos(8x)\right]\log(\cot x) \ .
\end{align}
However, for even $d$ there is a simpler method to solve equation \eqref{phijneq}, without the calculation of the integrals in \eqref{phinjint}. In that case the solution $\phi^{(n)}_j$ can be sought of as a sum of finite numbers of $p_n$ functions, as it was done in \cite{maliborski1}. While the method in \cite{maliborski1} works only in case of even spatial dimensions, our expansion can be performed for arbitrary $d$.

If the first solution $u_1$ is regular and localized then the other solution $u_2$ is necessarily singular at the center $x=0$ and non-localized at infinity $x=\pi/2$. In this case, assuming that the source term $F^{(n)}_j$ behaves well in \eqref{phijneq}, the second term in the inhomogeneous solution \eqref{phinjint} is always regular and localized, but the first term in \eqref{phinjint} can be regular and localized only if the integral in it vanishes both at $x=0$ and $x=\pi/2$. The choice $x_1=0$ makes the integral zero at the center, but in order to make it vanishing at infinity we have to require that
\begin{equation}
\int_{0}^{\pi/2}\frac{u_1}{W}F^{(n)}_j\mathrm{d}x=0 \ . \label{intcond}
\end{equation}
This gives a condition on $F^{(n)}_j$ in each case when $u_1$ is regular and localized. For example, for $j=1$ at $\varepsilon^2$ order $F^{(2)}_1$ is given by \eqref{f21eq}, and the condition \eqref{intcond} yields
\begin{equation}
0=\int_{0}^{\pi/2}\frac{u_1}{W}F^{(2)}_1\mathrm{d}x=
-2d\omega^{(1)}\int_{0}^{\pi/2}\cos^{d+1}x\sin^{d-1}x\,\mathrm{d}x
=-\frac{\sqrt{\pi}\,\Gamma\left(d/2+1\right)}{2^{d-1}\Gamma\left(d/2+1/2\right)}\omega^{(1)} \ ,
\end{equation}
from which follows that $\omega^{(1)}=0$. For each positive integer $n$, at $\varepsilon^{2n}$ order in the small-amplitude expansion the same source terms arise for $j=1$ as in \eqref{f21eq}, with $\omega^{(1)}$ replaced by $\omega^{(2n-1)}$, from which follows that
\begin{equation}
\omega^{(k)}=0 \ \ \text{for odd}\  k\ .
\end{equation}
After this, at $\varepsilon^{2n}$ order all $F^{(2n)}_j$ source terms are zero, and we take the trivial solution $\phi^{(2n)}_j=0$ for all Fourier components. Then $\phi^{(2n)}=0$, and as we have mentioned earlier, only odd powers of $\varepsilon$ remain in the expansion \eqref{epsexpansion} of $\phi$.

\subsection{Third order in the small-amplitude expansion}

To $\varepsilon^3$ order there are no scalar field source terms in the equations determining $A$ and $\delta$, hence only the trivial solution is regular, $A^{(3)}_j=\delta^{(3)}_j=0$. The same holds to all odd orders in $\varepsilon$. The only nonzero source terms in the scalar field equations \eqref{phijneq} at $\varepsilon^3$ order are
\begin{align}
F^{(3)}_1&=d^2\cos^dx\left[\left(\frac{2}{d}+2-\frac{1}{\cos^2x}\right)A^{(2)}_0
-2\omega^{(2)}-\frac{\cos^{2d}x}{4}\left(d-1+\cos(2x)+\frac{d}{\cos^2x}\right)\right] , \\
F^{(3)}_3&=\cos^{3d-2}x\left[-\frac{d^3}{4}+\frac{d^2}{4}(3d+2)\cos^2x
-\frac{d^2(3d+1)}{2(d+1)}\cos^4x\right] .
\end{align}
The integral condition \eqref{intcond} automatically holds for $F^{(3)}_3$, but for $F^{(3)}_1$ the condition determines the value of $\omega^{(2)}$. The integral of the term containing $A^{(2)}_0$ can be calculated by integration by parts using \eqref{a20eq2}, resulting in
\begin{equation}
\omega^{(2)}=-\frac{(3d+1)\sqrt{\pi}\;\Gamma\left(\frac{3d}{2}+1\right)}
{2^{2(d+1)}\Gamma\left(\frac{d}{2}\right)\Gamma(d+\frac{3}{2})} \ .
\end{equation}
In Table \ref{om2table} the value of $\omega^{(2)}$ is listed for several values of $d$.
\renewcommand\arraystretch{2}
\begin{table}[htbp]
\begin{tabular}{|c|c|c|c|c|c|c|c|c|}
\hline
$d$ & $2$ & $3$ & $4$ & $5$ & $6$ & $7$ & $8$ & $9$ \\
\hline
$\omega^{(2)}$ & $\displaystyle-\frac{7}{20}$ & $\displaystyle-\frac{45}{128}$
& $\displaystyle-\frac{13}{42}$ & $\displaystyle-\frac{65}{256}$
& $\displaystyle-\frac{57}{286}$ & $\displaystyle-\frac{24871}{163840}$
& $\displaystyle-\frac{25}{221}$ & $\displaystyle-\frac{21735}{262144}$\\
\hline
\end{tabular}
\caption{\label{om2table}
The value of $\omega^{(2)}$ for various spatial dimensions $d$.
}
\end{table}
The values of $\phi^{(3)}_1$ and $\phi^{(3)}_3$ can be calculated relatively easily by the integral formula \eqref{phinjint} for any concrete choice of $d$, but we could not find a simple expression which is valid for arbitrary dimensions.

\subsection{Higher order results for $d=3$} \label{sechighord}

For $d=3$ spatial dimensions we obtain
\begin{align}
\phi^{(3)}_1=&-\frac{27}{32}x^2\cos^3x
+\frac{27}{64}x\frac{\cos^{4}x}{\sin x}\left(\cos(2x)-3\right) \notag \\
&+\frac{3\cos^{3}x}{2895872}\left(743109-43632\cos(2x)-1212\cos(4x)-6464\cos(6x)+303\cos(8x)\right)
, \label{eqphi31}
\end{align}
and
\begin{equation}
\phi^{(3)}_3=\frac{3\cos^{3}x}{413696}\left(6845+4008\cos(2x)+7692\cos(4x)
-1368\cos(6x)+303\cos(8x)\right) . \label{eqphi33}
\end{equation}
The solutions of the differential equations determining $\phi^{(3)}_1$ and $\phi^{(3)}_3$ are not unique, since to any solution we can add a constant times the solution of the homogeneous equation, which is regular and localized in this case. Namely, one can replace $\phi^{(3)}_1$ by $\phi^{(3)}_1+c_1 p_1$, and $\phi^{(3)}_3$ by $\phi^{(3)}_3+c_3 p_3$, where $c_1$ and $c_3$ are some constants, and $p_n$ are the functions defined in \eqref{pneqgen}. The constant $c_1$ can be expressed by $c_3$ if we require that at $x=0$ we have $\phi^{(3)}_1+\phi^{(3)}_3=0$, which ensures that $\varepsilon$ really gives the central amplitude of $\phi$ at $\tilde{t}=0$ to $\varepsilon^3$ order in the expansion. The constant $c_3$ can still be arbitrary if we perform the expansion only up to $\varepsilon^3$ order. However, at $\varepsilon^5$ order it turns out that the equation for $\phi^{(5)}_3$ can have a regular localized solution only if the constant $c_3$ takes a certain value. The expressions \eqref{eqphi31} and \eqref{eqphi33} already give the corrected functions, for which $c_1=c_3=0$. However, we note that it is not possible to guess the proper functions just by solving the $\varepsilon^3$ order equations. The solution \eqref{eqphi33} does not correspond to a trivial choice of the integration constant $x_2$ in \eqref{phinjint}, such as $0$ or $\pi/2$.

At $\varepsilon^5$ order, from the condition of the existence of an appropriate solution for $\phi^{(5)}_1$, we get the fourth order correction to the frequency, which for $d=3$ takes the value
\begin{equation}
\omega^{(4)}=\frac{405}{1024}\pi^2-\frac{1208433249}{324337664}\approx 0.177656 \ .
\end{equation}

\begin{figure}[!hbtp]
\includegraphics[width=8cm,clip]{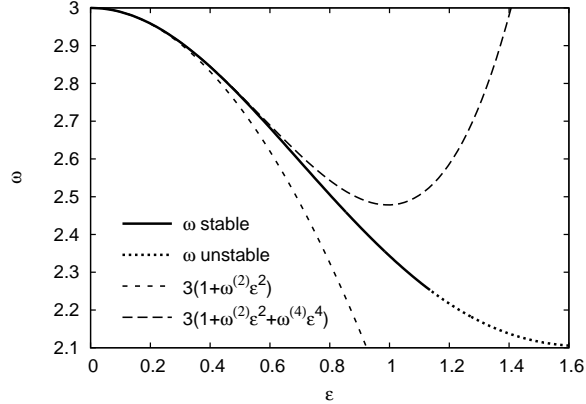}
\caption{\label{f:epsom} The oscillation frequency $\omega$ is shown as a function of the expansion parameter $\varepsilon$. For comparison, the small-amplitude expansion results, valid to $\varepsilon^2$ and $\varepsilon^4$ order are also given. According to the time-evolution code, the configurations with frequency below $\omega_s=2.253$ are unstable, so they are represented by a dotted line in that region.}
\end{figure}

On Fig.~\ref{f:epsom} we compare the small-amplitude expansion results to the precise results calculated by the spectral numerical code, by showing the frequency $\omega$ as the function of $\varepsilon$. The expansion gives reasonably good results in almost the whole range where the configurations are stable according to the time-evolution code. With the convention $8\pi G=d-1$ used in this section, $\varepsilon$ is equivalent to the central value of $\phi$ at the moment of time reversal symmetry. However, when comparing to the results in Sec.~\ref{sec:periodic}, one should rescale the scalar field with $2\sqrt{\pi}$, since the spectral numerical code was implemented with the choice $G=1$.

\begin{figure}[!hbtp]
\includegraphics[width=8cm,clip]{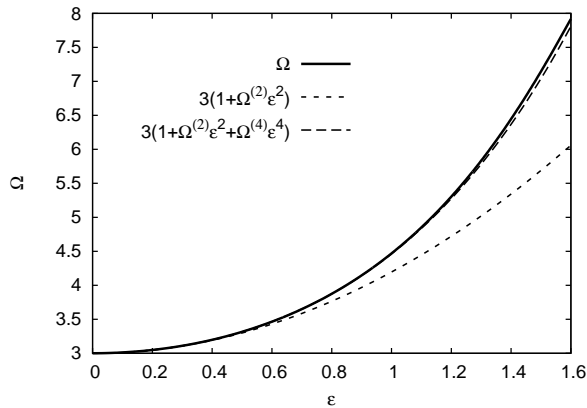}
\caption{\label{f:epsoom} Central frequency $\Omega$ as a function of the parameter $\varepsilon$. The solid line shows our numerical results, while the dashed lines give the small-amplitude expansion results.}
\end{figure}

Another interesting physical quantity is the frequency $\Omega$ as observed by a central observer. This can be calculated by integrating the proper time at the center during one oscillation period. From the numerical results of the spectral code this can be obtained as $\Omega=\omega/N_0(r=0)$. Using the small-amplitude formalism one has to use the central values of $\delta^{(2)}_0$, $\delta^{(2)}_2$ and $\delta^{(4)}_0$, and the expansion in $\varepsilon$ results in
\begin{equation}
\Omega=\omega^{(0)}\left(1+\Omega^{(2)}\varepsilon^2+\Omega^{(4)}\varepsilon^4+\ldots\right) .
\end{equation}
Here $\Omega^{(2)}=\omega^{(2)}+d/4$, which for $d=3$ spatial dimensions give $\Omega^{(2)}=51/128$. For $d=3$
\begin{equation}
\Omega^{(4)}=\frac{405}{8192}\pi^2-\frac{129387681}{324337664}\approx 0.398438 \ .
\end{equation}
On Fig.~\ref{f:epsoom} we compare the precise numerical value of $\Omega$ with the results obtained from the small-amplitude expansion.
The agreement with the $\varepsilon^4$ order result is remarkably good in all the domain for which we could compute configurations with the spectral numerical code. The likely reason for this good agreement is that both $\varepsilon$ and $\Omega$ are taken at the center of symmetry.

\begin{figure}[!hbtp]
\includegraphics[width=8cm,clip]{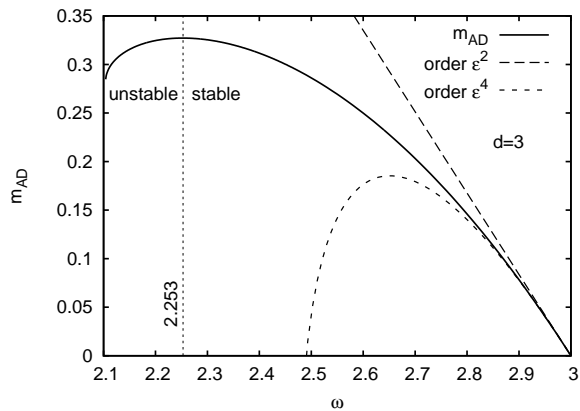}
\caption{\label{f:momega} Mass of the configurations as a function of their frequency for $d=3$ spatial dimensions. The $\varepsilon^2$ and $\varepsilon^4$ order results of the small-amplitude expansion are also shown for comparison. }
\end{figure}

The mass of the configurations can be calculated using the expansion \eqref{aexpans}. Only the time independent $A^{(n)}_0$ terms give contribution to the mass, and the result can be expanded as
\begin{equation}
m_{AD}=m_2\varepsilon^2+m_4\varepsilon^4+\ldots \  .
\end{equation}
For all dimensions, $m_2=9\pi/32$, and for $d=3$ we get
\begin{equation}
m_4=-\frac{81}{2048}\pi^3+\frac{638469}{11583488}\pi\approx -1.05316 \ .
\end{equation}
On Fig.~\ref{f:momega} we compare the precise numerical results to the first two nontrivial orders of the small-amplitude expansion.

It is quite remarkable that the perturbative expansion in the amplitude show excellent agreement with the numerical results up to $\varepsilon\approx 1/2$. We recall that $\varepsilon$ determines the value of the amplitude of the field at the center at $t=0$, therefore it is a physically meaningful quantity. Based on this, we conjecture that the perturbative expansion has a large ($\mathcal{O}(1)$) radius of convergence.

\section{Conclusions}

In this paper we have constructed spherically symmetric breathers in a massless scalar field theory coupled to Einstein's gravity in $d$ spatial dimensions imposing anti de Sitter asymptotics on space-time.  Using a time evolution code we have established stability for families of solutions of AdS breathers. We have worked out a perturbative constuction of the AdS breathers in the small-amplitude limit for any $d$, and we conjecture that it has a finite radius of convergence.

\begin{acknowledgments}

This research has been supported by OTKA Grant No. K 101709 and by the Marie Curie Actions Intra European Fellowship  of the European Community's Seventh Framework Programme under contract number PIEF-GA-2013-621992.
\end{acknowledgments}

\end{document}